\documentclass{article}
\RequirePackage{rotating}
\usepackage{float}
\usepackage{graphicx,epstopdf}
\usepackage{bbm}
\usepackage{theorem}
\usepackage{amsfonts}
\usepackage{booktabs}
\usepackage{multirow}
\usepackage{siunitx}
\usepackage{mathrsfs}
\usepackage{fancyhdr}
\usepackage{amsmath}
\usepackage{flexisym}
\usepackage{dsfont}
\usepackage{amssymb}
\usepackage{theorem}
\usepackage{graphicx}
\usepackage[dvips]{epsfig}
\usepackage{latexsym}
\usepackage{exscale}
\usepackage[latin1]{inputenc}
\usepackage{csquotes}
\usepackage{color}
\usepackage{rotating}
\usepackage{hyperref}
\def\ud{\, \mathrm{d}}
\numberwithin{equation}{section}

\newtheorem{theorem}{Theorem}[section]
\newtheorem{proposition}[theorem]{Proposition}
\newtheorem{lemma}[theorem]{Lemma}
\newtheorem{corollary}[theorem]{Corollary}
\newtheorem{remark}[theorem]{Remark}

\begin{document}
\title{A General Class of Multifractional Processes and Stock Price Informativeness}


\author{Qidi Peng, Ran Zhao}
\date{}
\maketitle



\begin{abstract}
We introduce a general class of stochastic processes driven by a multifractional Brownian motion (mBm) and study the estimation problems of their pointwise H\"older exponents (PHE) based on a new localized generalized quadratic variation approach (LGQV). By comparing our suggested approach with the other two existing benchmark estimation approaches (classic GQV and oscillation approach) through a simulation study, we show that our estimator has better performance in the case where the observed process is some unknown bivariate function of time and mBm. Such multifractional processes, whose PHEs are time-varying, can be used to model stock prices under various market conditions, that are both time-dependent and region-dependent. As an application to finance, an empirical study on modeling cross-listed stocks provides new evidence that the equity path's roughness varies via time and the stock price informativeness properties from global stock markets. 

\begin{flushleft}

\end{flushleft}
\end{abstract}




\section{Introduction}

Being a natural extension of Brownian motion (Bm) and fractional Brownian motion (fBm, see \cite{Mandelbrot1968}), multifractional Brownian motion (mBm) has nowadays been successfully applied to many fields such as finance, network traffic, biology, geology and signal processing, etc. Unlike Bm and fBm, mBm is a continuous-time Gaussian process whose increment processes are generally not stationary. However, the feature that multifractional process allows its local H\"older regularity to change via time makes the process flexible enough to model a much larger class of empirical data than the fBm does.

In literature, there exist several slightly different ways to define an mBm (see e.g. \cite{Benassi1997,Peltier1995,Ayache2004,Stoev2006}). In this paper we define an mBm $\{ X(t)\} _{t\in [0,1]}$ through the so-called harmonizable representation (see \cite{Benassi1997,Ayache2004}): for the time index $t\in[0,1]$,
\begin{equation}
\label{mBm}
X(t)=\int_{\mathbb R}\frac{e^{it\xi}-1}{|\xi|^{H(t)+1/2}}{\,\mathrm{d}} \widetilde{W}(\xi),
\end{equation} 
where:
\begin{description}
  \item[$-$] $H$ is called the pointwise H\"older exponent (PHE) of the $\{X(t)\}_{t\in[0,1]}$. 
      Recall that, for a continuous nowhere differentiable process $\{Y(t)\}_t$, its local H\"older regularity can be measured by the PHE. The PHE of $\rho_Y$ is a stochastic process defined by: for each $t_0$,
$$
\rho _{Y}(t_0)=\sup\bigg\{\alpha\in[0,1]:~\limsup\limits_{\varepsilon\rightarrow0}\frac{|Y(t_0+\epsilon)-Y(t_0)|}{|\varepsilon|^{\alpha}}=0\bigg\}.
$$
For the mBm $\{X(t)\}_t$,  it is shown by the zero-one law (see e.g. \cite{Ayache2004}) that its PHE $H$ is almost surely deterministic.
  \item[$-$] The complex-valued stochastic measure ${\,\mathrm{d}}\widetilde{W}$ is defined by the Fourier transform of the real-valued Brownian measure ${\,\mathrm{d}} W$. More precisely, for all $f$ belonging to the class of squared integrable functions over $\mathbb R$ (i.e. $f\in L^2(\mathbb R)$), we have
      $$
      \int_{\mathbb R}\widehat{f}(t){\,\mathrm{d}}\widetilde{W}(t)=\int_{\mathbb R}f(t){\,\mathrm{d}} W(t),
      $$
      where $\widehat{f}$ denotes the Fourier transform of $f$:
      $$
      \widehat{f}(\xi)=\int_{\mathbb R}e^{-i\xi t}f(t){\,\mathrm{d}} t,~\mbox{for all $\xi\in\mathbb R$}.
      $$
\end{description}
Multifractional processes, in particular mBm, come into vogue recently and are widely applied to financial modeling under empirical market conditions. For example, the last systemic financial crisis dated from $2007$ to $2009$ has strongly questioned the well-posedness of the classic dichotomy between efficient and inefficient markets. It is believed that the real financial markets are a complex system such that Bm and fBm are too reductive to explain it \cite{Bianchi2008}. Unlike fBm, mBm is flexible enough to overcome this inconvenience, mainly because its PHE can vary via time. Through an empirical study by Bianchi et al. \cite{Bianchi2008}, it was shown that the real-world stock prices can be modeled based on an mBm. Later, by estimating  the PHE of the stock price dynamics, Bianchi et al. \cite{Bianchi2013} find that the PHE fluctuates around $1/2$ (the sole value consistent with the absence of arbitrage), with significant deviations. In $2012$, Bertrand et al. \cite{Bertrand2012} introduce sparse modeling for mBm and apply it to NASDAQ time series. Recently, Bianchi et al. \cite{Bianchi2017} have suggested a new way to quantify how far from efficiency a market is at any fixed time $t$. Their dynamical approach, based on estimation of the time-varying PHE of the log-variations of the 3 stock indexes - Dow Jones Industrial Average (DJIA), the Dax (GDAXI) and the Nikkei 225 (N225), allows to detect the periods in which the market itself is efficient, once a confidence interval is fixed. Note that it is more difficult to estimate the mBm's PHE than the fBm's, due to the non-stationarity of the mBm's increment processes. This problem becomes even more challenging when modeling an \textit{individual} stock price (e.g. stock price of a particular entity) in lieu of averaged equity indexes, because the former one is not necessarily non-arbitrage and its corresponding PHE may be time-dependent and may take arbitrary values between 0 and 1. So far, there is not yet a satisfying model fitting the individual stock price process using multifractional processes. In this paper we aim to provide suitable models to describe these individual stock prices. Our main contribution consists of the following.
\begin{enumerate}
\item We introduce a general class of multifractional processes, that  can be used to describe the behavior of individual stock returns on equity markets. The proposed model is based on the assumption that the stock return is some unknown function of both time $t$ and an mBm at that time $t$ (see Section \ref{model}). 

\item Under the above assumption, we develop a new efficient approach to estimate the above model's PHE. The estimators from our approach are shown to be consistent (see Section \ref{sec:gqv_setup}).

\item  In Section \ref{sec:sim}, through a simulation study, we compare the performances of three estimation approaches (our new localized generalized quadratic variation approach (LGQV), the classic GQV approach and the oscillation method) on various functions $\Phi$.

\item In the empirical study (Section \ref{sec:app}), We apply the general multifractional process to model the individual stocks and use LGQV approach to estimate cross-listed stocks' PHEs. Then we determine the market factors that drive the individual stock returns' PHEs. The estimators of the PHEs reveal that the PHEs of individual stock prices are time-varying under various market conditions and their behaviors vary via different market regions. This interesting result enables us to examine the main individual stock's PHE drivers.
\end{enumerate}
Note that the Matlab codes used in Sections \ref{sec:sim} and \ref{sec:app} are provided. We conclude the paper in Section \ref{sec:con} and provide proof of the main result in Sections \ref{Proof}. The supplementary graphs, tables and other detailed technical proofs are given in Appendix see (Section \ref{Appendix}).

\section{A general class of multifractional processes}
\label{model}
Before introducing the general multifractional model that we are interested in, we briefly review the estimation of the multifractional process' PHE. 

In the multifractional process modeling problem, there is an obstacle : the PHE is basically not straightforwardly observed. The issue of estimating the PHE effectively arises. There are so far a number of estimation strategies existing in literature. We refer to \cite{Coeurjolly2005,Coeurjolly2006,Bertrand2013,Bardet2013,Jin2016,Trujillo2010} and the references therein.

Coeurjolly \cite{Coeurjolly2005,Coeurjolly2006} estimates the PHE of an mBm, starting from an observed discrete sample path of that mBm, using the LGQV approach (see also \cite{Chan1995}). Bertrand et al. \cite{Bertrand2013} study the same estimation problem as in \cite{Coeurjolly2005,Coeurjolly2006}, using the nonparametric estimation approach - increment ratio (IR) statistic method. This IR estimator has been later improved by Bardet and Surgailis \cite{Bardet2013} to the so-called pseudo-increment ratio approach, and it is applied to estimate the PHE of a more general multifractional Gaussian process (whose increments are asymptotically a multiple of an fBm) than mBm. There exist other approaches to estimate the PHE of fBm, that can be possibly extended to estimate the PHE of mBm. For example, in chaos theory and time series analysis, the statistical self-affinity is another measurement of the process path roughness. Since this exponent is tightly related to the PHE of self-similar processes (e.g. fBm), the detrended fluctuation analysis (DFA) methods developed by Peng et al. \cite{Peng19941,Peng19942} can be used to estimate the PHE of fBm. The time-varying PHE of mBm can be then approximated by applying the DFA piecewisely over time. However, the statistical self-affinity  is not equivalent to the PHE of a process, because it does not share all the properties of the Hausdorff dimension \cite{Peng19941,Peng19942}, while the Hausdorff dimension is equivalent to the PHE when the corresponding process is self-similar. In literature, it has been shown that the wavelet-based method is actually more accurate than the DFA on estimation of the PHE. Muzy et al. \cite{Muzy1991} have obtained representations of turbulence data and Brownian signals via wavelet decompositions. Bardet et al. \cite{Bardet2000} have applied the wavelet coefficient methods to estimate the PHE of long-memory processes (e.g. fBm with its PHE being greater than $1/2$), where some rate of convergence of the estimators are derived. Wendt et al. \cite{Wendt2009} have developed the wavelet leader based multifractal analysis for estimating 2D functions (images). Inspired by the above works, Jin et al. \cite{Jin2016} have provided a wavelet-based estimator of the time-varying PHE of a class of multifractional processes with a fine convergence rate, when the observations are the wavelet coefficients of some unknown function of a multiple of mBm, i.e. the observed process is of the form $\Phi(\theta(t)X(t))$, with $\Phi$ and $\theta$ being unknown $C^2$-functions. In both \cite{Bardet2013} and \cite{Jin2016}, estimators of PHE with fine convergence rates are constructed and strategies for selecting input parameters are discussed. 

\begin{center}
\begin{figure}[ht]
\label{fig:1}
\centering
\includegraphics[scale = 0.6]{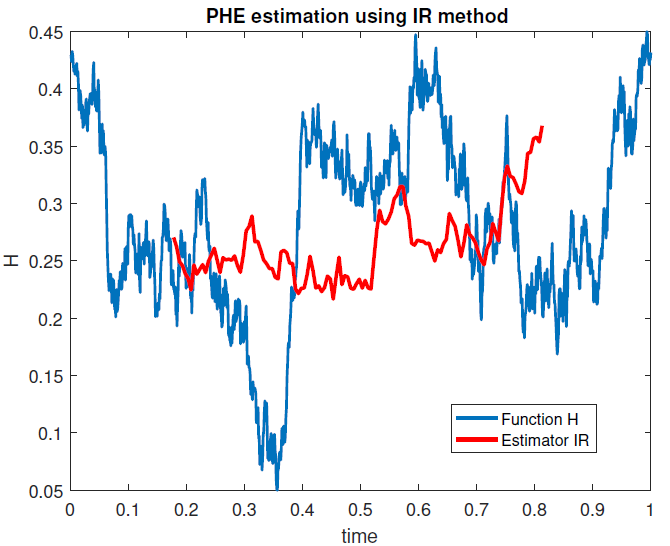}
\newline
\caption{\normalsize The graph illustrates the incremental ratio (IR) estimation \cite{Bardet2013} of the PHE for an mBm path. The blue line represents the true PHE, and the red line plots the IR estimated PHE. One of the drawbacks of the IR method is the loss of estimated PHE on the two-sided tails (starting and ending points).}
\label{fig::ir_est}
\end{figure}
\end{center}

Note that in our paper we also consider a model more general than the one in \cite{Jin2016}, in that it allows $\Phi$ to be a function of both $t$ and $x$, i.e. we assume the observed signal is some unknown function of time $t$ and mBm $X$: $\Phi(t,X(t))$. We apply the LGQV-based approach to estimate the PHE of $\Phi(t,X(t))$, when one of its discrete paths is observed. Similar to Jin et al. \cite{Jin2016}, an estimator with fine convergence rate is constructed and appropriate parameter selection is discussed. In \cite{Trujillo2010,Trujillo2012} the oscillation estimation method, which could be applied to estimate the PHE of all processes with continuous paths, is discussed. The main advantages of our approaches are: (1) The model is simple and general enough for finance application. (2) Compared to the oscillation estimation method, the LGQV method has higher accuracy and it allows us to select the input parameter from a large range of values. We will provide a fine rate of convergence of our LGQV estimator, which will further help practitioners to determine the best input parameter values. (3) One disadvantage of the increment ratio approaches is that, it is unable to estimate the PHE over the whole time interval $[0,1]$. Figure \ref{fig:1} is an example showing that, only part of the path of $\{H(t)\}_{t\in[0,1]}$ is estimated by the increment ratio method. However, the algorithm for LGQV-based approach can estimate $H$ pointwisely from $t=0$ to $t=1$. Moreover, it can be easily implemented using various programming languages such as Matlab, R and Python, etc.
Throughout this paper we consider the following model: for $t\in[0,1]$,
\begin{equation}
\label{Model1}
Z(t)=\Phi(t,X(t)),
\end{equation}
where
\begin{description}
\item[$-$] $\{X(t)\}_{t\ge0}$ is an mBm defined in (\ref{mBm}). Assume that its PHE $H$ belongs to the class of functions $C^{2}([0,1])$ (this means that $H$ is second-order continuously differentiable over $[0,1]$) and $[H_*,H^*]\subset(0,1)$, where $H_*=\inf_{t\in[0,1]}H(t)$ and $H^*=\sup_{t\in[0,1]}H(t)$.
  \item[$-$]  $\Phi$ is supposed to be an unknown deterministic $C^2(\mathbb R_+\times \mathbb R)$-function. Also we assume that $\partial_y \Phi(x,y)\neq 0$ for almost every $(x,y)\in\mathbb R_+\times\mathbb R\backslash\{0\}$ and there exist two constants $c_1,c_2$ such that  $0<c_1\leq |\partial_y\Phi(x,y)|\leq c_2$ for almost every $(x,y)\in\mathbb R_+\times\mathbb R\backslash\{0\}$.
  \item[$-$] Suppose that a discrete sample path of $Z$: $\big\{Z(u/2n):u=0,\ldots,2n\big\}$ is observed for some $n\in\mathbb N$ large enough.
\end{description}
From (\ref{Model1}) we see that the model $\{Z(t)\}_t$ is driven only by the time index $t$ and the mBm $\{X(t)\}_t$. It is quite general because the function $\Phi$ \enquote{lives} in a large class of functions  $C^2([0,1])$, and more importantly, it is supposed to be unknown. Examples of $\{Z(t)\}_t$ include the mBm, the self-regulating processes based on mBm \cite{Barriere2012}, and those in Section \ref{sec:sim}.

\section{LGQV estimation of the PHE}
\label{sec:gqv_setup}

Observing the discrete sample path of $Z(t)=\Phi(t,X(t))$:
$$
\left\{Z(0),\,Z\Big(\frac{1}{2n}\Big ),\,Z\Big(\frac{2}{2n}\Big ),\ldots,\,Z\Big(\frac{2n-1}{2n}\Big ),\,Z(1)\right\},
$$
where $n$ denotes an integer large enough, our goal is to propose a method allowing to estimate the PHE $H(t_0)$ of the hidden mBm $\{X(t)\}_{t\in [0,1]}$ at an arbitrary time $t_0\in (0,1)$. To this end, we apply a localized generalized quadratic
variations (LGQV) estimation method.
Before stating our main results, we need to briefly introduce some notations which will be used throughout the rest of the paper.
\begin{itemize}
\item As usual, $a=(a_0,\ldots, a_p)\in\mathbb R^{p+1}$ is
an arbitrary but fixed finite sequence having $Q\ge1$
vanishing moments, that is,
\begin{equation}
\label{moment:a:}
\sum_{k=0}^{p}k^{l}a_{k}=0, \mbox{for $l=0,\ldots,Q-1$ and } \sum_{k=0}^{p}k^{Q}a_{k}\neq 0.
\end{equation}
\item For all integer $n\ge p+1$ and $i\in\{0,\ldots,n-p-1\}$ the generalized increments of $Z$, $\Delta_{a}{Z}_{i,n}$, and that of $X$, $\Delta_{a}{X}_{i,n}$, are defined by,
\begin{equation}
\label{incYbarantch5}
  \Delta_{a}{Z}_{i,n}:=\sum_{k=0}^{p}a_{k}{Z}_{i+k,n}~\mbox{and}~ \Delta_{a}{X}_{i,n}:=\sum_{k=0}^{p}a_{k}{X}_{i+k,n},
\end{equation}
where ${Z}_{i+k,n}$ and ${X}_{i+k,n}$ denote the values of the processes $Z$ and $X$, at time $(i+k)/n$, that is,
\begin{equation}
\label{defYbarantch5}
{Z}_{i+k,n}:=Z\Big(\frac{i+k}{n}\Big)~\mbox{and}~{X}_{i+k,n}:=X\Big(\frac{i+k}{n}\Big).
\end{equation}
\item For all integer $n\ge p+1$, we denote by $\nu_{n}(t_0)$ the set of indexes defined by,
\begin{equation}
\label{eq9:ant-ch5}
\nu_{n}(t_0)=\bigg\{i\in\{0,\ldots,n-p-1\}:\,\,\Big|\frac{i}{n}-t_0\Big|\leq
v(n)\bigg\},
\end{equation}
where $v(\cdot)$ is an arbitrary function of $n\ge p+1$, valued in $(0,1]$, which satisfies for
each integer $n\ge p+1$, $v(n)\ge n^{-1}$ and $\lim_{n\rightarrow\infty} nv(n)=\infty$. Note that $\nu_{n}(t_0)$ labels the times $t$ in the neighborhood of $t_0$. As mentioned in \cite{Coeurjolly2005,Coeurjolly2006}, the estimation of $H(t_0)$ only relies on the observations of $Z(t)$, for $t$ being \enquote{neighbored to} $t_0$. Consequently, the estimation accuracy will be in terms of the size of the neighborhood selected for each $t_0$. Both theoretical and empirical studies tend to show that, the size of the neighborhood shouldn't be chosen too large nor too small, i.e. there is a trade-off between the estimator's rate of convergence and bias.
\item For all integer $n\ge p+1$ we define $n_{t_0}$ to be the number of points in $\nu_{n}(t_0)$:
\begin{equation}
\label{eq4:ant-ch5}
n_{t_0}=\#\nu_{n}(t_0).
\end{equation}
We then quickly observe that
\begin{equation}
\label{eq5:ant-ch5}
n_{t_0}\in \big\{[2n v(n)],[2n v(n)]+1\},
\end{equation}
where $[\cdot]$ is the integer part function.

\end{itemize}
The main results are presented in the next section.

\section{Estimation of the PHE}
\label{sec:estimYbarantch5}
In this section we construct a LGQV consistent estimator of $H(t_0)$, $t_0\in[0,1]$ of $\{Z(t)\}_{t\in[0,1]}$. Recall that, in statistics theory,  an estimator $\hat\theta_n$ of a parameter $\theta$ is (weakly) consistent if $\hat\theta_n$ converges to $\theta$ in probability, as $n\to\infty$. Let $\{U_n\}_{n\ge1}$ be an arbitrary sequence of random variables and $\{v_n\}_{n\ge1}$ be a sequence of non-vanishing real numbers, we use the notations
$$
U_n=\mathcal O_{a.s.}(v_n)~\mbox{and}~U_n=\mathcal O_{\mathbb P}(v_n)
$$
to denote
$$
\mathbb P\left(\sup_{n\ge1}\frac{|U_n|}{|v_n|}<\infty\right)=1~\mbox{and}~
\lim_{\eta\to\infty}\sup_{n\ge1}\mathbb P\left(\frac{|U_n|}{|v_n|}>\eta\right)=0,~\mbox{respectively}.
$$
Remark that $U_n=\mathcal O_{a.s.}(v_n)$ leads to $U_n=\mathcal O_{\mathbb P}(v_n)$, and the almost sure convergence implies the convergence in probability.

Our first main result below provides a delicate identification of the covariance of the generalized increments of the multifractional process of a particular form $Y(t)=\sigma(t)X(t)$. Later we will need this result for estimating the PHE of the more general function of $X(t)$: $Z(t)=\Phi(t,X(t))$.
\begin{proposition}
\label{maj:cov} Let $\{X(t)\}_{t\in[0,1]}$ be an mBm and $Y(t)=\sigma(t)X(t)$,
with $\sigma$ being a second order stochastic process independent of $X$, and the bivariate function $\theta(s,t):=\mathbb E(\sigma(s)\sigma(t))$ satisfies  $\theta\in C^2([0,1]^2)$. For a sequence $a\in\mathbb R^{p+1}$, assume $Q\ge 2$, then for every $k,k'\in\{0,1,\ldots,n-p-1\}$,  we have,
\begin{description}
\item[(1)] if $k\ne k'$,
\begin{eqnarray}
\label{kneqk'}
&&Cov(\Delta_a{Y}_{k,n},\Delta_a{Y}_{k',n})\nonumber\\
&&=C\left(H(\frac{k}{n})+H(\frac{k'}{n}),Q\right)\bigg(\frac{n^{-H(k/n)-H(k'/n)}\theta(k/n,k'/n)}{|k-k'|^{2Q-H(k/n)-H(k'/n)}}\bigg)\nonumber\\
&&\hspace{3cm}+n^{-H(k/n)-H(k'/n)}R(n,k,k',Q),
\end{eqnarray}
where
\begin{description}
\item[-] $C\in C^2([0,1]\times\mathbb R_+)$ is a deterministic function whose analytical expression is given in (\ref{def:C}) in Appendix.
\item[-] The remaining term $R(n,k,k',Q)$ satisfies
$$
\sum_{0\le k,k'\le n}|R(n,k,k',Q)|^2=\mathcal O\left(n^{-1}(\log(n))^2\right).
$$
\end{description}
\item[(2)] The variance of $\Delta_a{Y}_{k,n}$ can be identified as below:
\begin{equation}
\label{k=k'}
Var(\Delta_a{Y}_{k,n})=C_1\big(\frac{k}{n}\big)n^{-2H(k/n)}+\mathcal O(n^{-2H(k/n)-1}|\log n|^4),
\end{equation}
where the coefficient $C_1(k/n)$ is described in (\ref{def:C1}) in Appendix.
\end{description}
\end{proposition}
We quickly point out that, in the above proposition, there is not any inconvenience to regard $\sigma$ to be a $C^2([0,1])$ class deterministic function. It is also worth noting that Proposition \ref{maj:cov} has its own interests, since it gives an exact estimation of the covariance structure of the generalized increments of $Y(t)=\sigma(t)X(t)$, with a fine rate of convergence for its remaining term. To motivate the above facts we briefly compare Proposition \ref{maj:cov} to Lemma 1 in \cite{Coeurjolly2005} below:
\begin{enumerate}
\item Proposition \ref{maj:cov} extends Lemma 1 in \cite{Coeurjolly2005} from mBm $X(t)$ to a stochastic volatility process $Y(t)=\sigma(t)X(t)$, which is a useful model in financial time series analysis.
\item Unlike $(iii)$ in  Lemma 1 in \cite{Coeurjolly2005}, Proposition \ref{maj:cov} provides a finer identification of $\pi_H^a(k)$, by discovering a fine identification of the remaining term.
\item The only difference of assumptions on $H$ between Proposition \ref{maj:cov} and Lemma 1 in \cite{Coeurjolly2005} is that we assume $H\in C^2([0,1])$, while in \cite{Coeurjolly2005} it is assumed that $H\in C^{\eta}([0,1])$ with $\eta\in(0,1)$.
\end{enumerate}
The proof of Proposition \ref{maj:cov} is technical and long. It is moved to Section \ref{Appendix}: Appendix.

The second main result involves estimation of the PHE of the general process $Z(t)=\Phi(t,X(t))$, under a very general condition:
\begin{theorem}
\label{th:V1}  Pick a sequence $a\in\mathbb R^{p+1}$ with its first $Q\ge2$ moments being vanishing. We list the following conditions on $v(n)$.
\begin{itemize}
\item [(i)] $v(n)$ satisfies:
 $$
\lim_{n\to\infty}\sum_{l=0}^4v(n)^ln^{(l-2)H(t_0)}|\log n|^{2-l/2}=0,~\mbox{for all $t_0\in(0,1)$}.
$$
\item [(ii)] $v(n)$ satisfies:
 $$
\sum_{n\ge p+1}\frac{1}{(nv(n))^{2}}<\infty.
$$
\end{itemize}
For $t_0\in(0,1)$, define
\begin{equation}
\label{eq1:th:Vch5}
V_{n}(t_0)=\sum_{i\in\nu_{n}(t_0)}\big(\Delta_a{Z}_{i,n}\big)^2,
\end{equation}
and
\begin{equation}
\label{eq2:th:Vch5}
\widehat{H}_{n,t_0}=\frac{1}{2}\bigg(1+\log_2\Big(\frac{v(2n)}{v(n)}\Big)+\log_2\Big(\frac{V_{n}(t_0)}{V_{2n}(t_0)}\Big)\bigg),
\end{equation}
where $\log_2$ is the base-$2$ logarithm.\\
\textit{\textbf{(1)}} If $v(n)$ satisfies the condition $(i)$, we have
\begin{eqnarray*}
&&\widehat{H}_{n,t_0}-H(t_0)\nonumber\\
&&=\mathcal O_{a.s.}\left(\bigg(\sum_{l=0}^4v(n)^ln^{(l-2)H(t_0)}|\log n|^{2-l/2}\bigg)^{1/2}+v(n)^{H(t_0)}|\log(v(n))|^{1/2}\right)\nonumber\\
&&~~~~+\mathcal O_{\mathbb P}\left(v(n)\log n+v(n)^{-1}n^{-1}\right).
\end{eqnarray*}
\textit{\textbf{(2)}} If $v(n)$ satisfies the conditions $(i)$-$(ii)$, then
$$\widehat{H}_{n,t_0}\xrightarrow[n\rightarrow\infty]{a.s.}H(t_0),$$
where $\xrightarrow[n\rightarrow\infty]{a.s.}$ denotes the convergence almost surely.
\end{theorem}
Theorem \ref{th:V1} is our key result. It provides consistent estimators of the PHE of $\{Z(t)\}_t$ in a very general setting. Moreover, Theorem \ref{th:V1} (1) elaborates the rate of convergence of the estimators. We see that the rate of convergence depends only on the sample size $n$ and $v(n)$. The proof of Theorem \ref{th:V1} is provided in Section \ref{Proof}.
\section{Simulation study: selection of parameters and comparison with benchmark approaches} \label{sec:sim}

We conduct simulation studies to compare the performance of our PHE estimator provided in Theorem~\ref{th:V1} with the other two benchmark methods: the so-called classic GQV (see \cite{Barriere2007,Ayache2004}) and oscillation method (see \cite{Barriere2007}). Below we briefly introduce the classic GQV and oscillation method. 
\begin{itemize}
\item The GQV method  is applied to estimate the PHE of the so-called generalized mBm \cite{Ayache2004} and some multifractional signals \cite{Barriere2007}, it is based on the following result (see Theorem 2.2 in \cite{Ayache2004}):
    $$
    \lim_{n\to\infty}\frac{1}{2}\left(1-\gamma-\frac{\log V_{n}(t_0)}{\log n}\right)=H(t),
    $$
    where $\gamma=-\log v(n)/n$. Note that the above convergence holds almost surely if $H^*<\gamma<1/2$, however, milder condition exists for the convergence in probability to hold. We also remark that although our approach is also based on the classic GQV,  it is more complex than the above one, since in our case some constant $c_0$ should be introduced to cancel the unknown factor $\partial_y\Phi(t_0,X(t_0))$ in (\ref{eq1rem2:antch10}).
    \item The oscillation method is quite general and based on the following (see \cite{Trujillo2010} and the coding algorithm in \cite{Barriere2007}):
    $$
    H(t_0)=\liminf_{\varepsilon\to0}\frac{\log |Z(t_0+\epsilon)-Z(t_0)|}{\log|\varepsilon|}.
    $$
\end{itemize}
The implementation in Matlab codes of the above two approaches can be found in \textit{FracLab} by INRIA: \url{https://project.inria.fr/fraclab/}. We use the code version FracLab 2.2 in the empirical study\footnote{\url{https://project.inria.fr/fraclab/download/overview/}.}.

\subsection{Parameter selection}

To be more convenient, we denote by LGQV (Localized Generalized Quadratic Variation Method) the PHE estimation provided in Theorem~\ref{th:V1}. In this section, we discuss of the best choice of the functional parameter $v(n)$ in Theorem~\ref{th:V1}. In LGQV and classic GQV, we set the estimation neighborhood radius $v(n)$ to be of the form
$$
v(n)=n^{-\gamma(n)},~\mbox{with $\gamma(n)\in(0,1)$, which may depend on $n$}.
$$
In LGQV, to choose some $v(n)$ that satisfies the condition $(ii)$ in Theorem~\ref{th:V1}, we consider the following condition which is slightly stronger than the condition $(ii)$:
$$
v(n)^2n|\log n|^{4/3}=1,~\mbox{equivalently},~
\gamma(n) = \frac{1}{2} + \frac{2}{3}\frac{\log(\log(n))}{\log(n)}.
$$
The reason why we don't choose $\gamma\equiv 1/2$ but a stronger condition is to avoid the system computational error, which often has a strong impact when the parameter is close to an extreme value.

\begin{center}
\begin{figure}[pt]
\centering
\includegraphics[scale = 0.27]{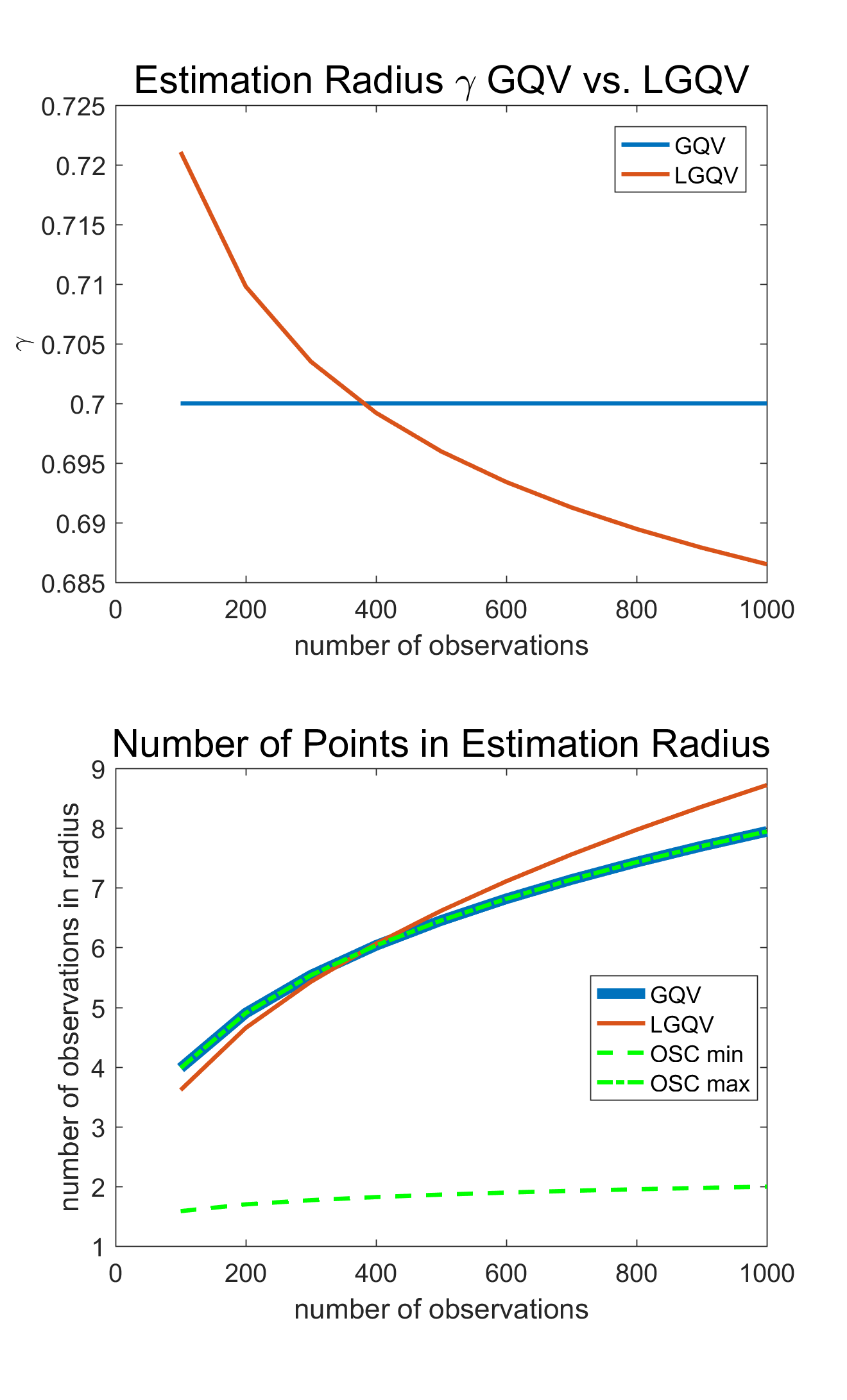}\newline
\caption{\normalsize The estimation radius parameter $\gamma(n)$ for the classic GQV and LGQV method is compared with increasing number of observations $n$ in the left graph. The right graph plots the number of points used for estimation corresponding to the number of observations $n$ for each method.}
\label{fig::est_nei}
\end{figure}
\end{center}

 As a benchmark, in the example of the classic GQV in \cite{Barriere2007}, the neighborhood radius parameter $\gamma$ is selected to be a constant $0.7$ throughout the simulation study. A noticeable feature of radius selection in LGQV is that $\gamma(n)$ decreases as $n$ increases, in contrast to the constant $\gamma$ selected by the classic GQV method. Moreover by this choice the neighborhood radius $v(n)$ of LGQV method is smaller than that of the classic GQV approach once $n\geq380$, as illustrated by the graph on the left hand-side  of Figure~\ref{fig::est_nei}.

In the oscillation method, a neighborhood for estimating the PHE at each point is required. By default we set $[n^{\alpha}, n^{\beta}]$ with $\alpha = 0.1$ and $\beta = 0.3$ in the approach (see \cite{Barriere2007,Trujillo2010,Trujillo2012}).

We have provided the theoretical choice on number of points in estimation neighborhood, which allows fraction number of points. In actual estimation using GQV and LGQV methods,  we round the theoretical number of points in neighborhood to its closest integer value. For oscillation method, the lower bound of neighborhood is rounded to ceiling integer value, whereas the upper bound of neighborhood is rounded to floor integer value.
\subsection{Experiments design}
In this section, we demonstrate simulation experiments that examine and compare estimation accuracy of LGQV estimator with conventional benchmark methods, i.e., the classic GQV with $\gamma\equiv0.7$ and oscillation methods, in various functional forms of $\Phi(t,X(t))$.

The PHE that we choose is $H(t) = 0.5 + 0.3\sin(2\pi t)$, for $t=0,1/n,\ldots,(n-1)/n,1$. We examine the convergence performance of estimators by using the simulated scenarios of $\{X(i/n)\}_{i=0,\ldots,n}$, with  $n$ ranging from 100 to 1000. We utilize Wood-Chan method~\cite{WoodChan} to generate independent scenarios of mBm. The scenarios of $\{\Phi(t,X(t))\}_t$ with various forms of $\Phi$, can therefore be generated. We then estimate the PHE straightforwardly starting from these scenarios of $\{\Phi(t,X(t))\}_t$.

We then compare the estimation outcome of LGQV with benchmark GQV and oscillation method, using scenarios from different categories of functional forms of $\Phi(t,X(t))$. The root-mean-squared error (RMSE) is used to measure and quantify PHE estimation performance. We use $100$ independent scenarios to compute each RMSE from the original $H(t)$.

We propose three categories of functional forms of $\Phi$. The first category contains single variate functions $\Phi(X(t))$ of mBm, where $\Phi$ is a $C^{\infty}(\mathbb R)$ function. With $\Phi$ being smooth enough, the transformed mBm in this category does not behave significantly differently from an mBm itself when the values of $X(t)$ are close to 0. Therefore, we expect that LGQV has similar estimation performance  to the benchmark methods. The specific functional forms in the first category include:
\begin{itemize}
\item $\Phi(t,X(t)) = X^2(t)$,
\item $\Phi(t,X(t)) = \exp(X(t))$.
\end{itemize}

The second category contains functional forms of both time and mBm, e.g. $\Phi(t,X(t))$. This functional type of mBm is where LGQV has advantage over the classic GQV method. It is important to remark that it is more reasonable to use this functional form $\Phi(t,X(t))$ to model financial time series. Note that the most commonly used form in financial derivative pricing is $\Phi(t,W(t))$ (where $W(t)$ denotes a Bm), on which an It\^o formula is often applied. Then our model $\Phi(t,X(t))$ naturally extends the latter one. We expect that the RMSE of LGQV is smaller in this category than the two benchmark approaches. The functional forms we take in this category include:
\begin{itemize}
\item $\Phi(t,X(t)) = \sin(t)X(t)$,
\item $\Phi(t,X(t)) = \sin^2(t) + X^2(t)$.
\end{itemize}

The third category of the functional form is $\Phi(\sigma(t),X(t))$, where $\sigma(t)$ is a stochastic process independent of $\{X(t)\}_t$ (this includes the case for $\sigma(t)$ being a deterministic function with continuous non-differentiable path). We would consider the case where $\sigma(t)=g(W(t))$ with $g$ being a smooth function. This is the scenario where both LGQV and classic GQV are not applicable. The PHE estimation of this type of functional form will be of interest for future research. The functional forms we consider in this category include:
\begin{itemize}
\item $\Phi(W(t),X(t)) = W(t)X(t)$,
\item $\Phi(W(t),X(t)) = W^2(t) + X^2(t)$,
\end{itemize}
where $\{W(t)\}_t$ is a Brownian motion independent of the mBm $\{X(t)\}_t$.

We perform simulation study with sample path lengths $n=100, 200, \ldots, 1000$. For each sample path length $n$, we simulate 100 independent scenarios of each of the above functional type and estimate its PHE using LGQV, the classic GQV and oscillation method, respectively. The averaged RMSE of the PHE estimation is calculated for each method. The speed of convergences of LGQV method with various differencing orders ($Q=2,3,4,5$) are illustrated and compared as well.

\subsection{Simulation results}
Before presenting the simulation results, it is interesting to consider the simplest functional form of mBm: $\Phi(t,X(t)) = X(t)$. Theoretically and intuitively, in this case the classic GQV method should have better converging performance than LGQV. The reason is that the latter approach estimates the first and second order terms of the Taylor expansion of $\Phi$, which is not necessary in this particular case. As a result, the LGQV slows down the estimator's convergence speed. The performances of both approaches when $\Phi(t,X(t)) = X(t)$ are shown in Table~\ref{table:sim_results} and left-top graph in Figure~\ref{fig::four_rest_plots} in Appendix (see Section \ref{Appendix:figures}). In this special case, it is not surprising that the classic GQV method outperforms LGQV in terms of lower averaged RMSE and estimation standard deviation.

\begin{center}
\begin{figure}
\centering
\includegraphics[scale = 0.7]{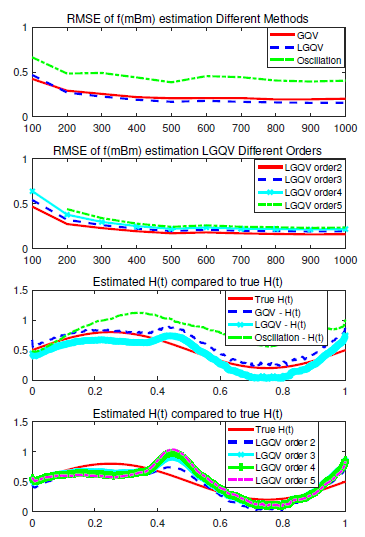}\newline
\caption{\normalsize The functional form is $\Phi(X(t))=\exp(X(t))$. The top graph compares the averaged RMSE (over 100 simulations) among the classic GQV (red solid), LGQV (blue dashed) and oscillation (green dotted dashed) methods over increasing number of points in each mBm path. The second top graph compares the averaged RMSE of various order of difference in LGQV method. The third and fourth graphs from the top draw the estimated $H(t)$ function by various methods on top of the true $H(t)$ used in simulating mBm. The true PHE is $H(t) = 0.5 + 0.3\sin(2\pi t)$.}
\label{fig::sim_t1_ep}
\end{figure}
\end{center}

In Figure~\ref{fig::sim_t1_ep}, we illustrate the comparison outcome of the PHE estimation for the first category functional form $\Phi(t,X(t))=\exp(X(t))$. From the figure we observe the following:
\begin{description}
\item[(1)] From the left-top graph, both the classic GQV and LGQV methods have consistently (averaged) converging tendency towards lower RMSE. The LGQV method outperforms the classic GQV method when $n\leq200$. And the averaged RMSE of LGQV method is $0.044$ ($21.65\%$), which is lower than that of the classic GQV method when $n=1000$.
    \item[(2)] The right-top graph shows that the averaged RMSE grows as the order of differences $p$ grows in the LGQV estimates. The differencing order of 2 has the lowest estimation error. The reason is that higher differencing order reduces the number of observed points for each estimation neighborhood, hence it reduces the accuracy of the estimation. High order generalized variations could perform better when the number of observations $n$ is relatively large.
        \item[(3)] The left bottom graph shows that the oscillation method overestimates $H(t)$.
        \item[(4)] In the two bottom graphs we see that the classic GQV and LGQV estimators are definitely better the oscillation estimator. This is because these two approaches are specifically designed for Gaussian processes. But they seem to have a ``hump'' shape delaying to the peak of true $H(t)$. The classic GQV method has worse delayed overestimation than LGQV method. The higher order the differencing is in LGQV method, the severe the overestimation following a peak on H\"older parameters. In addition, the LGQV method overreacts when the true $H(t)$ drops down, but has better performance on tail estimation (around the tail edge in this context). The good performance of the classic GQV comes from the reason that $\exp(x)$ is close to $x$ when $x$ is near 0, as in the case in our context.
\end{description}
\begin{center}
\begin{figure}
\centering
\includegraphics[scale = 0.9]{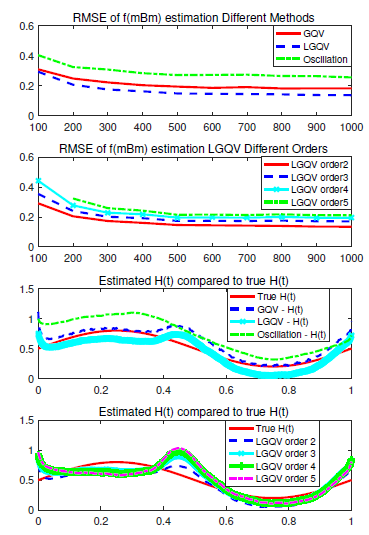}\newline
\caption{\normalsize The functional form is $\Phi(t,X(t))=\sin(t)X(t)$. The graphs have the same plotting mechanism (but different content) as that of Figure~\ref{fig::sim_t1_ep}, respectively.}
\label{fig::sim_t2_ep}
\end{figure}
\end{center}

In Figure~\ref{fig::sim_t2_ep}, we illustrate the comparison of PHE estimation results for $\Phi(t,X(t))=\sin(t)X(t)$. The outcome confirms the fact that LGQV consistently outperforms the classic GQV method in terms of lower RMSE, when analyzing the functional form $\Phi(t,X(t))$. Quantitatively, the averaged RMSE by LGQV (when $n=1000$) for this functional form is $0.135$, which is $0.045$ less than the classic GQV case. Both the classic GQV and LGQV methods outperform the oscillation method, again due to the classic GQV approaches' specialization on estimating PHE for Gaussian processes. The oscillation estimator of $H(\cdot)$ has similar shape to the its target, but it constantly overestimates $H(t)$ for each $t$ and has PHE estimation exceeding 1, due to the system error.
\begin{center}
\begin{figure}
\centering
\includegraphics[scale = 0.85]{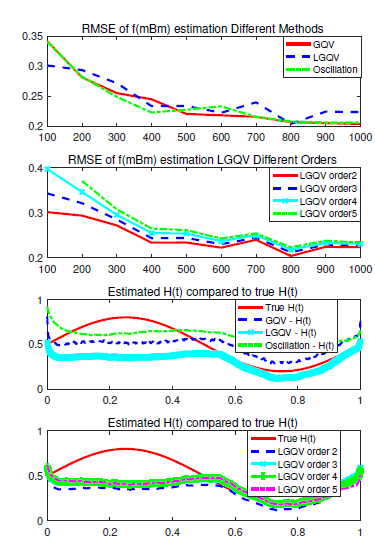}\newline
\caption{\normalsize The functional form of mBm is $\Phi(W(t),X(t))=W(t)X(t)$. The graphs have the same plotting mechanism (but different content) as that of Figure~\ref{fig::sim_t1_ep}, respectively.}
\label{fig::sim_t3_ep}
\end{figure}
\end{center}

Figure~\ref{fig::sim_t3_ep} shows the RMSE comparison of PHE estimation for different methods within the third functional form category, where neither the classic GQV nor LGQV estimation is applicable. The function considered is $\Phi(W(t),X(t))=W(t)X(t)$, where $W(t)$ is an $X(t)-$independent Brownian motion. The standard deviation of RMSE for LGQV method is higher than the ones for the competing methods. The partial reason for LGQV method not being applicable is that the assumption that $\partial_y \Phi(x,\cdot)\in C^2(\mathbb R)$ for $x\in\mathbb R_+$ is violated. Another reason that the classic GQV performs better is that the LGQV is more sensitive than the classic GQV and we have selected wider neighborhood radius than the classic GQV does in this simulation study (see Figure~\ref{fig::est_nei}).

More detailed PHE estimation comparison statistics of each method when $n=1000$ with 100 simulations are listed in Table~\ref{table:sim_results}. It shows the numerical comparison results for all functional forms considered in this simulation study. The corresponding figures are also given in Figure~\ref{fig::four_rest_plots} in Section \ref{Appendix}. The conclusion from these examples is consistent with what we have elaborated above.

\begin{table}[htb]
\tiny
\begin{center}
\caption{This table presents the averaged RMSE, standard deviation, maximum and minimum of each method settings when number of points in mBm path $n=1000$. }
\begin{tabular}{c|c|c|c|c|c|c|c}
  \hline
$\Phi(t,X(t))$  &  Stats. &  GQV  &  LGQV (2)  &  LGQV (3)  &  LGQV (4)  &  LGQV (5)  &  OSC  \\ \hline
	&	avg.	&	0.13068	&	0.1369	&	0.1742	&	0.1998	&	0.2187	&	0.1797	\\	
$X(t)$	&	std.	&	0.02061	&	0.0224	&	0.0279	&	0.0307	&	0.0324	&	0.0396	\\	
	&	max.	&	0.18558	&	0.1871	&	0.235	&	0.2666	&	0.2931	&	0.2759	\\	
	&	min.	&	0.07911	&	0.0787	&	0.0968	&	0.1147	&	0.1313	&	0.1076	\\	\hline
	&	avg.	&	0.1837	&	0.1394	&	0.1727	&	0.1955	&	0.2119	&	0.2504	\\
$X(t)^2$	&	std.	&	0.0263	&	0.0224	&	0.0269	&	0.0298	&	0.0321	&	0.077	\\
	&	max.	&	0.252	&	0.1894	&	0.2448	&	0.2755	&	0.2912	&	0.5204	\\
	&	min.	&	0.1213	&	0.081	&	0.0924	&	0.1053	&	0.12	&	0.1147	\\ \hline
	&	avg.	&	0.2032	&	0.1592	&	0.1919	&	0.2157	&	0.2337	&	0.4043	\\
$\exp(X(t))$	&	std.	&	0.101	&	0.0899	&	0.0919	&	0.0971	&	0.1042	&	0.6456	\\
	&	max.	&	1.1086	&	0.9641	&	1.0001	&	1.0614	&	1.1371	&	6.3053	\\
	&	min.	&	0.122	&	0.0912	&	0.1075	&	0.1165	&	0.1244	&	0.1175	\\ \hline
	&	avg.	&	0.1804	&	0.1352	&	0.171	&	0.195	&	0.2119	&	0.2538	\\
$\sin(t)X(t)$	&	std.	&	0.0262	&	0.0214	&	0.0267	&	0.0305	&	0.033	&	0.0449	\\
	&	max.	&	0.2522	&	0.1826	&	0.2257	&	0.2574	&	0.278	&	0.4135	\\
	&	min.	&	0.1249	&	0.0767	&	0.0913	&	0.1034	&	0.1115	&	0.158	\\ \hline
	&	avg.	&	0.1838	&	0.1394	&	0.1727	&	0.1955	&	0.2119	&	0.2556	\\
$\sin(t)^2+X(t)^2$	&	std.	&	0.0263	&	0.0224	&	0.0269	&	0.0298	&	0.0321	&	0.0789	\\
	&	max.	&	0.252	&	0.1894	&	0.2448	&	0.2755	&	0.2912	&	0.5202	\\
	&	min.	&	0.1214	&	0.081	&	0.0924	&	0.1053	&	0.12	&	0.1163	\\ \hline
	&	avg.	&	0.2038	&	0.2234	&	0.2278	&	0.2309	&	0.2342	&	0.2064	\\
$W(t)X(t)$	&	std.	&	0.0286	&	0.0296	&	0.0329	&	0.0348	&	0.036	&	0.0381	\\
	&	max.	&	0.2817	&	0.2928	&	0.3034	&	0.3126	&	0.3165	&	0.2823	\\
	&	min.	&	0.1423	&	0.1716	&	0.1609	&	0.1544	&	0.1515	&	0.1212	\\ \hline
	&	avg.	&	0.2568	&	0.2432	&	0.2455	&	0.2482	&	0.2525	&	0.3169	\\
$W(t)^2+X(t)^2$	&	std.	&	0.0351	&	0.0349	&	0.0378	&	0.0397	&	0.0413	&	0.0543	\\
	&	max.	&	0.3541	&	0.3447	&	0.3595	&	0.365	&	0.3648	&	0.4659	\\
	&	min.	&	0.1752	&	0.1716	&	0.1611	&	0.1495	&	0.1447	&	0.1836	\\
  \hline
\end{tabular}
\label{table:sim_results}
\end{center}
\end{table}

We summarize the advantage of LGQV method over the conventional benchmark PHE estimation methods. The converging rate of LGQV method improves significantly, when the underlying process takes functional form of time and mBm. Specifically in our study, the averaged RMSE decreases by $22.88\%$ in the single variate functional $\Phi(X(t))$ cases and decreases by $24.61\%$ in the bivariate functional $\Phi(t,X(t))$ cases. The standard deviation of PHE estimation reduces by $12.91\%$ and $16.57\%$ in the two functional form cases, respectively. Therefore, the LGQV method is applicable to a much larger set of multifractional processes in practice: $C^2(\mathbb R_+\times\mathbb R)$ functional form of $(t,X(t))$.

\section{An empirical study: application to financial time series} \label{sec:app}
Recently, Keylock \cite{Keylock2018} has formulated the gradual multifractal reconstruction approach and applied it to stock market returns spanning the 2008 crisis. By comparing the relation between the normalized log-returns and their H\"older exponent for the daily returns of eight financial indexes, Keylock observes that the change for NASDAQ 100 and S$\&$P 500 from a non-significant to a strongly significant cross-correlation between the returns and their H\"older exponents from before the 2008 crash to afterwards. However Asian markets don't exhibit significant cross-correlation to those from elsewhere globally.

Our setting and goal in this section is different from \cite{Keylock2018}. We apply the functional form $\Phi(t,X(t))$ to model equity prices and examine the PHE of individual stock series using the proposed LGQV method, from three markets. Then we compare the PHEs of the above three markets. Note that other than equity indexes and stock prices, the current financial literature has also introduced mBm in modeling stochastic volatilities \cite{Corlay2014} and currency exchange rates \cite{Bianchi2012}, etc.

We improve the work of \cite{Bianchi2013} by exploring the functional form of mBm in equity model and extend \cite{Corlay2014} by providing a practical PHE estimation method using mBm framework. To our best knowledge, this framework is also the first one to provide a financial interpretation of the PHE: we interpret the PHE as a quantitative measure of stock price informativeness on equity market. That is, PHE can be regarded as a new dimension on measuring information content embedded in stock price series. To see this, we provide the motivation on describing equity price informativeness by PHE and establish the connection by showing the shared characteristics of these two concepts.

\subsection{PHE and price informativeness}
To motivate the interpretation, we summarize three common features of stock market informativeness and the PHE of price series.

\begin{center}
\begin{figure}[ht]
\centering
\includegraphics[scale = 0.36]{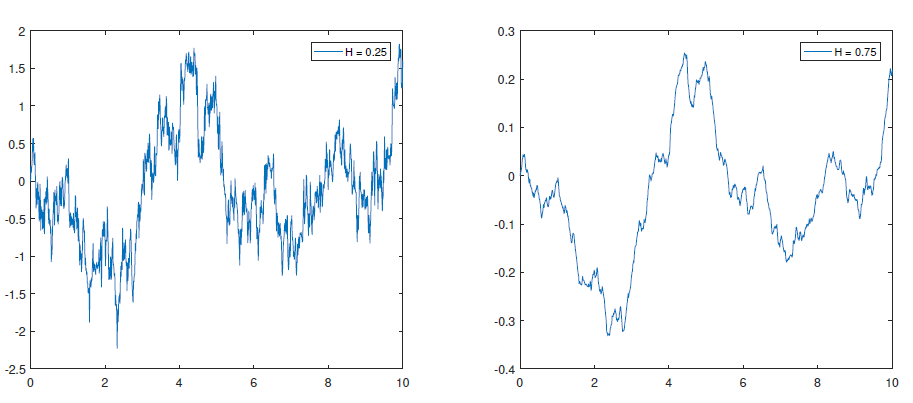}\newline
\caption{\normalsize The left graph plots a realized fBm path with $H\equiv0.25$, whereas the right graph plots a realized fBm path with $H\equiv0.75$.}
\label{fig::sim_comp_Ht}
\end{figure}
\end{center}

First note that, the more information is contained on the market, the lower is the PHE value of the stock price series, ceteris paribus. Consider the two simulated mBm paths plotted in Figure~\ref{fig::sim_comp_Ht} with different PHEs. With smaller value of its PHE (see the graph on the left-hand side), the mBm path is \enquote{rougher} and presents more local fluctuations within certain time interval than that of larger  PHE value case (see the graph on the right-hand side). These ``zig-zags'' in local ranges are analogous to arrivals of information to the market and investors' digest of the information. More intense arrivals of information generate more local fluctuations, resulting in a lower PHE value.

Secondly, the PHEs of price series and price informativeness of stocks are both time-varying. That is, the quantity of information on the market for the same underlying stock should not be a constant. The time-varying feature of information content orients from business cycle and cyclically financial reporting procedure regulated by SEC. In addition, corporate events, such as merger and acquisition, stock repurchase and executive management change etc., randomly introduce pertinent information to the equity market. The property of multifractionality reflects this time-varying feature of information flows, where our informativeness measure (PHE) $H$ allows to be a function of time. Bianchi et al. \cite{Bianchi2013} point out that the equity index has its PHEs close to $0.5$, but deviations are also observed over the past decades, which supports the non-constant PHE argument under the mBm framework.

Thirdly, corporate characteristics and market conditions influence the information content, as well as the PHE of the underlying stock. More specifically, corporate factors such as business complexity, growth stage, and idiosyncratic features may cause their stocks' PHEs to be significantly different with each other. For instance, a company with high analyst coverage is expected to have more information on the market and thus lower PHE. Market effects, such as capitalization size, accounting quality and efficiency, would also contribute to changing the PHE values  of its stock, from one period to another.

Given the informativeness context, we estimate and compare the PHEs of cross-listed stocks. The empirical study explores the main market factors that impact stock price informativeness.

\subsection{Other informativeness measures in literature}

Our informativeness measure provides new vision to common believes of information content carriers in the existing literature. Accounting and financial reports are conventionally regarded as the main information sources \cite{Huang14,CCDLS14}. Traditional measures of information content in finance literature also include quote revisions \cite{H91}, internet stock message boards \cite{AF04} and implied volatility from option markets \cite{DL92} etc. Besides corporate level effects, different markets and market conditions determine the information content as well. Morck et al. \cite{MYY00} suggest that stock markets in developed economics are more useful as information processor than the stock markets in emerging economics. They argue that better property rights protection in advanced economics explains the fact that stock prices synchronize more closely in emerging markets than in advanced economics. Ivkovi\'c and Weisbenner \cite{IW05} show that there is significant asymmetric information between geographical local and nonlocal investors. In addition, empirical study shows that local holdings generates excess return over nonlocal holdings. These findings seem to support the home bias theory \cite{CM99,VNV09}.

We propose the PHE to be a new quantitative measure of all the above price informativeness on equity markets. As one example, we examine whether this new measure is lieu with the existing stock price informativeness measures in the following two sections.

\subsection{Empirical study on real data}

This empirical study aims to disentangle the main factors that explain informativeness and informative flow on global equity markets. We focus on a special set of equities: cross-listed stocks. The cross-listed stock refers to as same underlying entity but its stocks are listed on multiple equity markets. The advantage of choosing cross-listings are twofold. Firstly, corporate characteristics keep the same when considering same underlying entity. Secondly, comparing price informativeness measure (PHE) provides evidence on revealing the real information source and answer the question  \cite{K06}: where does  price discovery occur?

We select cross-listed stocks that are listed on the following three equity markets simultaneously for more than 10 years:

 \begin{description}
 \item[(1)] China Mainland (SSE and SZSE, home market).
 \item[(2)] Hong Kong (SEHK).
 \item[(3)] The U.S. (NYSE) stock markets
 \end{description}

The three-market setting enables us to uniquely compare pairwisely the factors that determine informativeness. Empirical results will lead us to the conclusion of the debating question: \emph{which factor contributes more on information content, location or market quality?}

To be specific, China Mainland's market locates very close to Hong Kong markets in contrast to the far US market. The term ``location'' here includes not only geographical exchange location, but also language, time zone, culture and local knowledge etc., which is commonly believed to bring more information to home market investors than foreign market investors. Meanwhile, Hong Kong and the US stock markets are widely believed to be the ones with best market quality, in terms of their easier capital accessibility and sophisticated accounting system, where advanced financial and accounting information drives informativeness significantly.

The China Mainland and Hong Kong markets have location in common but different levels of market quality. Hence the comparison of PHE between cross-listed stock in China Mainland and HK markets reveal the effect of market quality on stock price informativeness. At the meantime, Hong Kong and the US markets have same level of market quality but different locations. Then the comparison of PHE between HK and US markets addresses the effect of home bias on information content.

Table~\ref{table::dual_list_stocks} in Appendix presents all cross-listed stocks that satisfy the condition on simultaneously listing on US, HK and China markets for more than 10 years. We exclude the mutual exclusive non-trading days for these three markets, and yield three time series with common trading days for each of the stock. Our empirical data contains 11 stocks and 33 price time series (for three markets in total). The estimation period starts from the earliest available date and ends up to December 30, 2016. Among the samples, the smallest number of observations for each market is $2261$, and the largest number of observations is $5558$.

We model the equity price series by $\Phi(t,X(t))$, an unknown functional form of time and mBm, and estimate each series' PHE as price informativeness measure. The empirical PHE comparison of the stocks is presented in the next subsection.

\begin{sidewaystable}[ph!]
\centering
\begin{center}
\caption{The table lists all the sample stocks in empirical study. The US, HK and China Mainland markets ticker (T. exchange name) and listing date (Date exchange name) for the stocks are provided. The company name is the registered institutional name on the US market. The daily US market stock prices and listing date are from CRSP database, daily HK stocks prices and their listing dates are from HKSE (Hong Kong Exchange) website (\url{https://www.hkex.com.hk/}), and China Mainland market information is from CSMAR database.}
{\normalsize
\resizebox{\columnwidth}{3cm}{%
\begin{tabular}{c|c|c|c|c|c|c|c}
  \hline
Company Name	&	T. NYSE	&	Sector	&	Date NYSE 	&	Date SHSE	&	Date SEHK 	&	T. SHSE 	&	T. SEHK	\\ \hline
Aluminum Corporation Of China	&	ACH	&	Primary Production of Aluminum	&	Dec. 12, 2001	&	April 30, 2007	&	Dec. 12, 2001	&	601600	&	2600	\\
China Eastern Airlines	&	CEA	&	Air Transportation	&	Feb. 4, 1997	&	Nov. 5, 1997	&	Feb. 5, 1997	&	600115	&	670	\\
China Unicom	&	CHU	&	Radiotelephone Communications	&	June 21, 2000	&	Oct. 9, 2002	&	June 22, 2000	&	600050	&	762	\\
Guangshen Railway	&	GSH	&	Railroads	&	May 13, 1996	&	Dec. 22, 2006	&	May 14, 1996	&	601333	&	525	\\
Huaneng Power International	&	HNP	&	Electric Services	&	Oct. 6, 1994	&	Dec. 6, 2001	&	Jan. 21, 1998	&	600011	&	902	\\
China Life Insurance	&	LFC	&	Life Insurance	&	Dec. 17, 2003	&	Jan. 9, 2007	&	Dec. 18, 2003	&	601628	&	2628	\\
PetroChina	&	PTR	&	Crude Petroleum \& Natural Gas	&	April 4, 2000	&	Nov. 5, 2007	&	April 7, 2000	&	601857	&	857	\\
Sinopec Shanghai Petrochemical	&	SHI	&	Crude Petroleum \& Natural Gas	&	July 26, 1993	&	Nov. 8, 1993	&	July 26, 1993	&	600688	&	338	\\
China Petroleum \& Chemical	&	SNP	&	Petroleum Refining	&	Oct. 18, 2000	&	Aug. 8, 2001	&	Oct. 19, 2000	&	600028	&	386	\\
Yanzhou Coal Mining	&	YZC	&	Bituminous Coal \& Lignite Surface Mining	&	Mar. 31, 1998	&	July 1, 1998	&	April 1, 1998	&	600188	&	1171	\\
China Southern Airlines	&	ZNH	&	Air Transportation	&	July 30, 1997	&	July 25, 2003	&	July 31, 1997	&	600029	&	1055	\\
  \hline
\end{tabular}}}
\label{table::dual_list_stocks}
\end{center}
\end{sidewaystable}


\subsection{Empirical results}
We separate the whole estimation period (earliest available up to December 30, 2016) into three subintervals: prior, within and post global financial crisis periods. In our study, the subprime crisis time frame is defined as from the beginning of year 2007 up till the end of 2009. The definition of financial crisis in our paper is longer than common definitions. But the information formation, propagation and digestion usually take longer time than the extreme movements observed on the stock market.

\begin{center}
\begin{figure}
\centering
\resizebox{1\textwidth}{!} {\includegraphics[scale = 0.05]{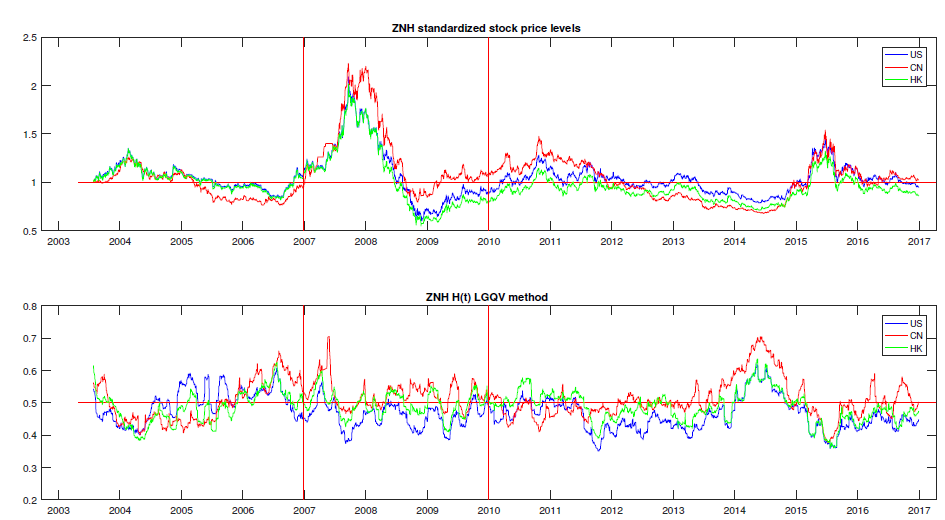}}
\newline
\caption{\normalsize The top graph plots the rescaled stock price levels of China Southern Airlines (NYSE: ZNH). The initial stock price sets to 1, and price levels follows the daily log-return of the stock. Stock levels above the horizontal reference line represent that the current stock price exceeds its initial value, and visa versa. The bottom graph illustrates the estimated PHE by LGQV method. The horizontal reference line in second graph is a benchmark PHE at 0.5, where the process is a standard Brownian motion. Two vertical reference lines in both graphs indicate the threshold of prior, within and post financial crisis time periods. The red line represents the dynamics of China market. Green line represents the dynamics of Hong Kong market. And the blue line stands for the dynamics in the US market.}
\label{fig::emp_ZNH}
\end{figure}
\end{center}

This separation of time periods allows us to analyze the change of PHE over different market performances. The first 100 trading observations are excluded to avoid the price instability in the post-IPO period. To illustrate our empirical findings, the rescaled stock price (initial price to be 1) and corresponding estimated PHE of China Southern Airlines (NYSE ticker: ZNH) are shown in Figure~\ref{fig::emp_ZNH}.

Observations from bottom graph in Figure~\ref{fig::emp_ZNH} prove that the PHE in single stock price series is not constant and may deviate from 0.5. This deviation highly depends on market performance. During the financial crisis period, the PHE is the lowest for all three markets. The US market has the most significant drop on PHE during financial crisis, which is also the orient market of the crisis. The PHE increases when the stock price becomes directional, as in the year 2014. Though the stock prices are highly positively correlated among the global markets, the PHE might be less positively correlated or even negatively correlated, as in the year 2005. This observation provides new evidence to the segmented market hypothesis: there are different driving factors of informativeness and stock prices for the same stock at various markets.

The relationship of PHEs among three markets also changed structurally from prior to post financial crisis periods. There are 8 out of 11 individual stocks in our sample that have prior financial crisis observations, where 4 stocks have lower PHEs in the US market than that of China Mainland and HK markets. The home bias mixes the effect of more sophisticated markets and market participants. However, in post financial crisis time period, 10 of 11 stocks have lower PHE in US market than that in China Mainland markets\footnote{We utilize two-sample $t$ test in mean to conclude significant difference of averaged PHEs between the two markets. The test significance level in our study is 5\%.}. In addition, there are 10 stocks whose US PHEs are lower than their HK counterparts. And 8 out of 11 stocks have lower HK market PHE than that of China Mainland market. Home bias seems to lose its dominance on price informativeness or opinion formation. Market sophistication, such as total capitalization, analyst coverage and developed accounting system etc., starts to determine information content on the financial market after the financial crisis. Another plausible explanation of low US market PHEs is that, the trading time (of the same trading day) of US market is running behind that of China Mainland and HK markets. The price informativeness on the Asian markets would increase the informativeness to the US market participants.

For robustness check, we use benchmark approaches to estimate PHE on the same stock price series. More details about the PHE statistics in various markets, time periods and estimation methods are presented in Table~\ref{table::stock_Ht}. The empirical findings on PHE comparison remain the same in the classic GQV and oscillation methods as that in LGQV method. Our conclusion seems to be valid cross various PHE estimation approaches.

\begin{table}[ph!]
\tiny
\begin{center}
\caption{The table below compares the estimated averaged H\"older exponent ($H(t)$) for each stock price series. The stock ticker (US market ticker) is on the leftmost column. The $H(t)$ of the stocks are estimated by the classic GQV, LGQV and oscillation method (column name GQV, LGQV and OSC respectively) for different markets. Prior, within and post are relative time periods to the most recent global financial crisis. Prior represents trading dates prior to year 2007, if available. Within represents trading dates within year 2007 and 2008, which is the duration of subprime mortgage crisis in our study. Post stands for post-crisis period, which include trading dates post to year 2008.}
\resizebox{\columnwidth}{6cm}{%
\begin{tabular}{c|c|ccc|ccc|ccc}
  \hline
Ticker	&	Market	&		&	Prior	&		&		&	Within	&		&		&	Post	&		\\ \cline{3-11}
	&		&	GQV	&	LGQV	&	OSC	&	GQV	&	LGQV	&	OSC	&	GQV	&	LGQV	&	OSC	\\ \hline
	&	CN	&	-	&	-	&	-	&	0.4898	&	0.4901	&	0.5087	&	0.4848	&	0.4594	&	0.6596	\\	
ACH	&	HK	&	-	&	-	&	-	&	0.5517	&	0.5045	&	0.4689	&	0.4857	&	0.4515	&	0.6204	\\	
	&	US	&	-	&	-	&	-	&	0.4776	&	0.4637	&	0.4823	&	0.4563	&	0.4255	&	0.6229	\\	\hline
	&	CN	&	0.5391	&	0.4663	&	0.5126	&	0.6187	&	0.5402	&	0.3872	&	0.5862	&	0.5056	&	0.5073	\\	
CEA	&	HK	&	0.5218	&	0.4597	&	0.4731	&	0.564	&	0.4901	&	0.3922	&	0.5453	&	0.4845	&	0.476	\\	
	&	US	&	0.588	&	0.5074	&	0.4887	&	0.4823	&	0.4444	&	0.3968	&	0.5038	&	0.4543	&	0.4992	\\	\hline
	&	CN	&	0.5041	&	0.4674	&	0.5581	&	0.4282	&	0.4363	&	0.4131	&	0.5154	&	0.4753	&	0.5158	\\	
CHU	&	HK	&	0.5038	&	0.4416	&	0.4528	&	0.4715	&	0.4107	&	0.371	&	0.4838	&	0.4496	&	0.4361	\\	
	&	US	&	0.4416	&	0.4214	&	0.4552	&	0.4487	&	0.4033	&	0.3717	&	0.4854	&	0.4508	&	0.4402	\\	\hline
	&	CN	&	0.4284	&	0.4149	&	0.4223	&	0.4211	&	0.4101	&	0.4806	&	0.5152	&	0.4643	&	0.5833	\\	
GSH	&	HK	&	0.5009	&	0.3961	&	0.461	&	0.4866	&	0.4298	&	0.4352	&	0.4706	&	0.4408	&	0.4941	\\	
	&	US	&	0.3197	&	0.3065	&	0.3977	&	0.4292	&	0.4006	&	0.4268	&	0.441	&	0.4183	&	0.493	\\	\hline
	&	CN	&	0.5318	&	0.5107	&	0.4682	&	0.5052	&	0.4606	&	0.417	&	0.5043	&	0.4744	&	0.5187	\\	
HNP	&	HK	&	0.4436	&	0.4189	&	0.4697	&	0.5031	&	0.4577	&	0.4035	&	0.5192	&	0.4692	&	0.474	\\	
	&	US	&	0.4656	&	0.4364	&	0.4686	&	0.4687	&	0.4338	&	0.4005	&	0.4823	&	0.4366	&	0.475	\\	\hline
	&	CN	&	-	&	-	&	-	&	0.4645	&	0.4458	&	0.4758	&	0.5034	&	0.4672	&	0.5826	\\	
LFC	&	HK	&	-	&	-	&	-	&	0.4488	&	0.4422	&	0.4228	&	0.5237	&	0.487	&	0.4986	\\	
	&	US	&	-	&	-	&	-	&	0.3829	&	0.3958	&	0.4393	&	0.4858	&	0.4613	&	0.5301	\\	\hline
	&	CN	&	-	&	-	&	-	&	0.4936	&	0.4923	&	0.5446	&	0.5192	&	0.473	&	0.6644	\\	
PTR	&	HK	&	-	&	-	&	-	&	0.493	&	0.4564	&	0.4473	&	0.5591	&	0.5065	&	0.5278	\\	
	&	US	&	-	&	-	&	-	&	0.4851	&	0.4507	&	0.4498	&	0.5178	&	0.4846	&	0.5271	\\	\hline
	&	CN	&	0.544	&	0.4866	&	0.5309	&	0.5657	&	0.4833	&	0.3806	&	0.603	&	0.5019	&	0.4333	\\	
SHI	&	HK	&	0.5533	&	0.4801	&	0.4306	&	0.4939	&	0.44	&	0.3886	&	0.4847	&	0.4209	&	0.428	\\	
	&	US	&	0.5501	&	0.4881	&	0.4448	&	0.4447	&	0.4106	&	0.3908	&	0.473	&	0.4111	&	0.4517	\\	\hline
	&	CN	&	0.523	&	0.4738	&	0.5876	&	0.4926	&	0.4687	&	0.4337	&	0.5278	&	0.4805	&	0.5837	\\	
SNP	&	HK	&	0.5352	&	0.4451	&	0.558	&	0.5027	&	0.4324	&	0.4356	&	0.5148	&	0.4561	&	0.5135	\\	
	&	US	&	0.5082	&	0.4406	&	0.569	&	0.4549	&	0.4035	&	0.4389	&	0.4882	&	0.433	&	0.5222	\\	\hline
	&	CN	&	0.5024	&	0.4543	&	0.4527	&	0.5201	&	0.4585	&	0.3357	&	0.5547	&	0.4935	&	0.4124	\\	
YZC	&	HK	&	0.5534	&	0.454	&	0.4872	&	0.5962	&	0.4962	&	0.3748	&	0.5729	&	0.5105	&	0.4642	\\	
	&	US	&	0.5566	&	0.4739	&	0.4529	&	0.5436	&	0.4665	&	0.3525	&	0.5087	&	0.4607	&	0.4523	\\	\hline
	&	CN	&	0.5444	&	0.4984	&	0.5812	&	0.5731	&	0.5259	&	0.4419	&	0.553	&	0.5102	&	0.5703	\\	
ZNH	&	HK	&	0.5512	&	0.4863	&	0.5519	&	0.5422	&	0.496	&	0.4678	&	0.5413	&	0.486	&	0.5484	\\	
	&	US	&	0.5564	&	0.4993	&	0.5588	&	0.4853	&	0.4584	&	0.4584	&	0.5108	&	0.4587	&	0.5296	\\	
  \hline
\end{tabular}}
\label{table::stock_Ht}
\end{center}
\end{table}

\section{Conclusion} \label{sec:con}

In this paper, we introduce a general multifractional process of the form $\Phi(t,X(t))$, driven by the mBm $X(t)$. We then propose a consistent estimator of PHE of this process. The estimation algorithm is based on identifying the localized quadratic variation statistic. A good convergence condition is found (see Theorem~\ref{th:V1}) and the selection of parameters in the algorithm is discussed.

In the simulation study we show that the proposed LGQV estimator outperforms the other two benchmark approaches in terms of lower PHE estimation RMSE and standard deviation, when the underlying process is an unknown $C^2$ function of time $t$ and the mBm $X(t)$.

In the empirical study, we show that the PHE can be interpreted as a quantitative measurement of the stock price informativeness. We use PHE estimated by LGQV to perform an empirical study regarding the impact of market quality and home bias on stock market behaviors. The comparison results indicate that in most recent years, the market quality dominates local knowledge and becomes the main driver of stock price informativeness.

Finally it is worth noting that, there exist other measurements to describe the local regularity of a process, such as the local H\"older exponent and the statistical self-affinity by the DFA methods. These exponents are generally not equivalent and no one could entirely capture the local regularity. A stochastic 2-microlocal analysis \cite{Herbin2009} is recently developed to provide a finer characterization of the local regularity. This approach particularly describes how the PHEs and the local H\"older exponents evolve subject to (pseudo-)differential operators and  multiplication by power functions. Another benefit from the 2-microlocal analysis is that it can be used to derive
local behaviour of sample paths from the regularity of the integrand and the integrator. Therefore a more general problem arises:  estimation of the local behavior of the stochastic integrals driven by a general class of multifractional processes. We leave this problem for future research.

\section{Proof of Theorem \ref{th:V1}}
\label{Proof}
The proof of Theorem \ref{th:V1} mainly relies on the identification of the localized generalized quadratic variation of mBm, and a bivariate Taylor expansion of the function $\Phi$. For identifying the  localized generalized quadratic variation of mBm we have the following proposition.
\begin{proposition}
\label{lemmeVN2}
Let $\widetilde{C}_a :(0,1)\rightarrow (0,+\infty)$ be the function defined by: for all $\alpha\in (0,1)$,
\begin{equation}
\label{eq1:prop1:ant-ch5}
\widetilde{C}_a (\alpha)=2\int_{\mathbb R} \frac{(1-\cos \eta)\big |\sum_{k=0}^p a_k e^{ik\eta}\big |^2}{|\eta|^{2\alpha+3}}\ud \eta.
\end{equation}
Then, assuming that $v(n)$ satisfies the condition $(i)$ in Theorem~\ref{th:V1}, we have,
$$
\frac{\sum_{i\in\nu_{n}(t_0)}\big(\Delta_a{X}_{i,n}\big)^2}{2\widetilde{C}_a (H(t_0))
v(n)n^{1-2H(t_0)}}=1+\mathcal O_{\mathbb P}\left(v(n)\log n+v(n)^{-1}n^{-1}\right).
$$
\end{proposition}
In order to prove Proposition~\ref{lemmeVN2}, we need several preliminary results, namely, Lemmas \ref{lem1:ant-ch5} - \ref{lem2:antch5} as follows.

Lemma \ref{lem1:ant-ch5} below is obtained based on the property that the moments of any order of a Gaussian random variable are equivalent. Its proof is quite similar to that of Lemma 6.3.5 in \cite{Peng2011'}, so we omit it.
\begin{lemma}
\label{lem1:ant-ch5}
For each $t_0\in(0,1)$, there is a constant $c>0$ such that for all integer $n\ge p+1$, we have
\begin{eqnarray}
\label{var4:}
&&\mathbb E\bigg(\sum_{i\in\nu_{n}(t_0)}(\Delta_a{X}_{i,n})^2-\sum_{i\in\nu_{n}(t_0)}Var(\Delta_a{X}_{i,n})\bigg)^4\nonumber\\
&&~~~~\le c\bigg(Var\Big(\sum_{i\in\nu_{n}(t_0)}(\Delta_a{X}_{i,n})^2\Big)\bigg)^2.
\end{eqnarray}
\end{lemma}

The following lemma (see Lemma 6.3.6 in \cite{Peng2011'} for its proof) results from a Gaussian vector's feature.
\begin{lemma}
\label{lemmegauss}
Let $(Z_1,Z_2)$ be an arbitrary zero-mean 2-dimensional Gaussian vector such that $Var(Z_1)=Var(Z_2)=\tau$. Then we have,
$$
\mathbb E\big((Z_1Z_2)^2-\tau^2\big)=2(Cov(Z_1,Z_2))^2.
$$
\end{lemma}

Next we state two consequences of Proposition \ref{maj:cov}. The result below can be derived from the proof of (\ref{k=k'}) in Proposition \ref{maj:cov} (it suffices to take $H$ to be a constant $H(t_0)$).
\begin{corollary}
\label{lem1:antch5}
Let $\{B_{H(t_0)}(s)\}_{s\in[0,1]}$ be the fBm with Hurst parameter $H(t_0)$. For any integer $n\ge p+1$ and any
$j\in\{0,\ldots,n-p-1\}$, define the following generalized increment:
\begin{equation}
\label{eq4:ant-idenB}
\Delta_a{B}_{j,n}^{H(t_0)}:=\sum_{k=0}^pa_k{B}_{j+k,n}^{H(t_0)}=\sum_{k=0}^pa_kB_{H(t_0)}\left(\frac{j+k}{n}\right).
\end{equation}
Then we have,
\begin{equation}
\label{eq1:lem1:antch5}
Var\left(\Delta_a{B}_{j,n}^{H(t_0)}\right)=\widetilde{C}_a (H(t_0)) n^{-2H(t_0)},
\end{equation}
where $\widetilde{C}_a$ is the function defined in (\ref{eq1:prop1:ant-ch5}).
\end{corollary}

Another straightforward consequence of Proposition \ref{maj:cov} is the following statement:
\begin{corollary}
\label{rem1:antch5}
There exist two constants $0<c'\le c$ such that for all $n$ big enough and all $j\in\nu_n (t_0)$, we have
\begin{equation}
\label{eq1:rem1:antch5}
c'n^{-2H(t_0)}\le Var(\Delta_a{X}_{j,n})\le cn^{-2H(t_0)}.
\end{equation}
\end{corollary}

Lemma \ref{iden:X} below aims to identify the localized generalized quadratic variation of the mBm in terms of its variances.
\begin{lemma}
\label{iden:X}
Assume that $v(n)$ satisfies the condition $(i)$ in Theorem~\ref{th:V1}. Then, we have
$$
\frac{\sum_{i\in\nu_{n}(t_0)}(\Delta_a{X}_{i,n})^2}{\sum_{i\in\nu_{n}(t_0)}Var(\Delta_a{X}_{i,n})}=1+\mathcal O_{\mathbb P}\left(v(n)^{-2}n^{-2}\right).
$$
\end{lemma}

The following results in Lemma \ref{lem2:antch5} are derived from the mBm's continuous paths property. More precisely, that property is, the mBm $\{X(t)\}_t$'s paths behave almost surely as an fBm's paths with Hurst parameter $H(t_0)$ when $t$ takes values in the neighborhood of $t_0$.
\begin{lemma}
\label{lem2:antch5}
Assume that $v(n)$ satisfies the condition $(ii)$ in Theorem~\ref{th:V1}. Then, we have,
\begin{equation}
\label{eq23:ant-idenB}
\max_{j\in\nu_n(t_0)}\Big\{E\Big |\Delta_a{X}_{j,n}-\Delta_a{B}_{j,n}^{H(t_0)}\Big|^2\Big\}=
\mathcal O \big ((v(n)\log (n))^2 n^{-2H(t_0)}\big);
\end{equation}
\begin{equation}
\label{ident1}
\max_{j\in \nu_{n}(t_0)}\bigg|\frac{Var(\Delta_a{X}_{j,n})}{Var(\Delta_a{B}_{j,n}^{H(t_0)})}-1\bigg|=
\mathcal O(\log(n)v(n));
\end{equation}
and
\begin{equation}
\label{eq1:lem3:antch5}
\bigg|\frac{\sum_{j\in\nu_n(t_0)}Var(\Delta_a{X}_{j,n})}{\sum_{j\in\nu_n(t_0)}Var(\Delta_a{B}_{j,n}^{H(t_0)})}-1\bigg|=
\mathcal O(\log(n)v(n)).
\end{equation}
\end{lemma}

Now we are ready to prove Proposition~\ref{lemmeVN2}.\\
\textbf{Proof of Proposition~\ref{lemmeVN2}.} First we can decompose $\frac{\sum_{i\in\nu_{n}(t_0)}(\Delta_a{X}_{i,n})^2}{2\widetilde{C}_a(H(t_0))v(n)n^{1-2H(t_0)}}$ into the product of 3 terms:
\begin{eqnarray}
\label{diffLemme4}
&&\frac{\sum_{i\in\nu_{n}(t_0)}(\Delta_a{X}_{i,n})^2}{2\widetilde{C}_a(H(t_0))v(n)n^{1-2H(t_0)}}=\bigg(\frac{\sum_{i\in\nu_{n}(t_0)}(\Delta_a{X}_{i,n})^2}{\sum_{i\in\nu_{n}(t_0)}Var(\Delta_a{X}_{i,n})}\bigg)\nonumber\\
&&\times\bigg(\frac{\sum_{i\in\nu_{n}(t_0)}Var(\Delta_a{X}_{i,n})}{\sum_{i\in\nu_{n}(t_0)}Var(\Delta_a{B}_{i,n}^{H(t_0)})}\bigg)
\bigg(\frac{\sum_{i\in\nu_{n}(t_0)}Var(\Delta_a{B}_{i,n}^{H(t_0)})}{2\widetilde{C}_a(H(t_0))v(n)n^{1-2H(t_0)}}\bigg).
\end{eqnarray}
Next we observe from (\ref{eq1:lem1:antch5}) in Corollary \ref{lem1:antch5} and (\ref{eq4:ant-ch5}) that,
\begin{equation}
\label{sumvB}
\sum_{i\in\nu_{n}(t_0)}Var(\Delta_a{B}_{i,n}^{H(t_0)})=n_{t_0}\widetilde{C}_a(H(t_0))n^{-2H(t_0)},
\end{equation}
and (\ref{eq5:ant-ch5}) entails that,
\begin{equation}
\label{cardN}
n_{t_0}=2nv(n)+\mathcal O(1).
\end{equation}
Thus, combining (\ref{sumvB}) with (\ref{cardN}), we get
\begin{equation}
\label{conv1}
\frac{\sum_{i\in\nu_{n}(t_0)}Var(\Delta_a{B}_{i,n}^{H(t_0)})}{2\widetilde{C}_a(H(t_0))v(n)n^{1-2H(t_0)}}=1+\mathcal O(v(n)^{-1}n^{-1}).
\end{equation}
Finally, Proposition~\ref{lemmeVN2} results from (\ref{diffLemme4}), Lemma
\ref{iden:X}, (\ref{eq1:lem3:antch5}) in Lemma \ref{lem2:antch5} and (\ref{conv1}):
\begin{eqnarray*}
&&\frac{\sum_{i\in\nu_{n}(t_0)}(\Delta_a{X}_{i,n})^2}{2\widetilde{C}_a(H(t_0))v(n)n^{1-2H(t_0)}}=\left(1+\mathcal O_{\mathbb P}(v(n)^{-2}n^{-2})\right)\left(1+\mathcal O(v(n)\log n)\right)\nonumber\\
&&~~\times\left(1+\mathcal O(v(n)^{-1}n^{-1})\right)=1+\mathcal O_{\mathbb P}\left(v(n)\log n+v(n)^{-1}n^{-1}\right).~\square
\end{eqnarray*}
In addition to Proposition \ref{lemmeVN2}, we also need the following Lemmas \ref{lemmevn3} and  \ref{lemmeV13} to prove the main result Theorem \ref{th:V1}.
\begin{lemma}
\label{lemmevn3} For any integer $n\geq p+1$, we define
\begin{equation}
\label{eq1:lemmevn3antch5}
V_{n,1}(t_0)=\Big(\partial_y\Phi(t_0,X(t_0))\Big)^2\sum_{i\in\nu_{n}(t_0)}(\Delta_a{X}_{i,n})^2,
\end{equation}
and
\begin{equation}
\label{eq2:lemmevn3antch5}
V_{n,2}(t_0)=\sum_{i\in\nu_{n}(t_0)}\Big(\partial_y\Phi\Big(t_0,X\big(\frac{i}{n}\big)\Big)\Big)^2 (\Delta_a{X}_{i,n})^2.
\end{equation}
Then,
\begin{equation}
\label{eq3:lemmevn3antch5}
\frac{V_{n,2}(t_0)}{V_{n,1}(t_0)} = 1+\mathcal O_{a.s.}\left(v(n)^{H(t_0)}|\log(v(n))|^{1/2}\right).
\end{equation}
\end{lemma}
\textbf{Proof of Lemma \ref{lemmevn3}.} By definitions (\ref{eq1:lemmevn3antch5}) and (\ref{eq2:lemmevn3antch5}), it is clear that
\begin{eqnarray}
\label{diffv341}
&&V_{n,2}(t_0)-V_{n,1}(t_0)\nonumber\\
&&=\sum_{i\in\nu_{n}(t_0)}\!\!\!\!\Big(\Big(\partial_y\Phi\Big(t_0,X\big(\frac{i}{n}\big)\Big)\Big)^2-\Big(\partial_y\Phi(t_0,X(t_0))\Big)^2\Big)(\Delta_a{X}_{i,n})^2.
\end{eqnarray}
By assumption there exists a random variable $C_1>0$ with all its moments being finite such that,
\begin{equation}
\label{fpol:antch5}
\sum_{l=0}^2 \sup_{u\in[-\|X\|_\infty,\|X\|_\infty]}\left|\partial_y^l\Phi(t_0,u)\right|\leq C_1,
\end{equation}
where $\|X\|_{\infty}=\sup_{s\in [0,1]} |X(s)|$ is a Gaussian random variable with its moments of any order being finite (see \cite{Rosenbaum2008,Ledoux2010}).
Using the mean value theorem and the inequality (\ref{fpol:antch5}), we obtain that, 
\begin{eqnarray}
\label{difff}
&&\left|\Big(\partial_y\Phi\Big(t_0,X\big(\frac{i}{n}\big)\Big)\Big)^2-\Big(\partial_y\Phi(t_0,X(t_0))\Big)^2\right|\nonumber\\
&&=\left|\partial_y\Phi\Big(t_0,X\big(\frac{i}{n}\big)\Big)+\partial_y\Phi(t_0,X(t_0))\right|\left|\partial_y\Phi\Big(t_0,X\big(\frac{i}{n}\big)\Big)-\partial_y\Phi(t_0,X(t_0))\right|\nonumber\\
&&\leq C_2 \left(\sup_{u\in[-\|X\|_{\infty},\|X\|_{\infty}]}\left|\partial_y^{2}\Phi(t_0,u)\right|\right)\left|X\big(\frac{i}{n}\big)-X(t_0)\right|\nonumber\\
&&\leq C_3\left|X\big(\frac{i}{n}\big)-X(t_0)\right|,
\end{eqnarray}
where $C_2=2C_1$ (given in (\ref{fpol:antch5})), $C_3=C_1C_2$. We now recall an important result on the path behavior of the mBm $\{X(s)\}_{s\in [0,1]}$ (see Theorem 1.7 in \cite{Benassi1997}); namely, there is a positive random variable $C_4$ of finite moments of any order, such that
for all $s,s'\in [0,1]$,
\begin{equation}
\label{diffX} |X(s)-X(s')|\leq
C_{4}|s-s'|^{\max\{H(s),H(s')\}}|\log|s-s'||^{1/2},~\mbox{a.s.}.
\end{equation}
As a particular case,
$$
|X(s)-X(s')|=0,~\mbox{when}~s=s'.
$$
It follows from (\ref{difff}), (\ref{diffX}) and the inequalities: there is a constant $c>0$ such that for all $i\in\nu_n(t_0)$,
\begin{equation}
\label{diffHt}
\left|H(\frac{i}{n})-H(t_0)\right|\le c\left|\frac{i}{n}-t_0\right|\le cv(n),
\end{equation}
that, for all $i\in\nu_n(t_0)$,
\begin{eqnarray}
\label{difff1}
&&\left|\Big(\partial_y\Phi\Big(t_0,X\big(\frac{i}{n}\big)\Big)\Big )^2-\left(\partial_y\Phi(t_0,X(t_0))\right)^2\right|\nonumber\\
&&\le C_3C_4\Big|\frac{i}{n}-t_0\Big|^{H(t_0)}\left|\log\Big|\frac{i}{n}-t_0\Big|\right|^{1/2}\left(\Big|\frac{i}{n}-t_0\Big|^{\max\{H(i/n),H(t_0)\}-H(t_0)}\right)\nonumber\\
&&\le C_3C_4\Big|\frac{i}{n}-t_0\Big|^{H(t_0)}|\log(v(n))|^{1/2}\left(v(n)^{-cv(n)}\right)\nonumber\\
&&\le C_5 v(n)^{H(t_0)}|\log(v(n))|^{1/2},
\end{eqnarray}
where $C_5=C_3 C_4\sup\limits_{n\in\mathbb N,n\ge p+1}v(n)^{-cv(n)}<\infty$. Finally, (\ref{eq1:lemmevn3antch5}), (\ref{eq2:lemmevn3antch5}), the triangle inequality, (\ref{diffv341}) and (\ref{difff1}), imply that
 \begin{eqnarray*}
 &&\left |\frac{V_{n,2}(t_0)}{V_{n,1}(t_0)}-1\right|=\left |\frac{\sum_{i\in\nu_{n}(t_0)}\big(\partial_y\Phi(t_0,X(i/n))\big)^2(\Delta_a{X}_{i,n})^2}{\big (\partial_y\Phi(t_0,X(t_0))\big)^2\sum_{i\in\nu_{n}(t_0)}(\Delta_a{X}_{i,n})^2}
-1\right |\\
&&\leq\frac{\sum_{i\in\nu_{n}(t_0)}\Big|\big(\partial_y\Phi(t_0,X(i/n))\big)^2-\big(\partial_y\Phi(t_0,X(t_0))\big)^2\Big|(\Delta_a{X}_{i,n})^2}{\big (\partial_y\Phi(t_0,X(t_0))\big)^2\sum_{i\in\nu_{n}(t_0)}(\Delta_a{X}_{i,n})^2}\\
&&\leq \frac{C v(n)^{H(t_0)}|\log(v(n))|^{1/2}\sum_{i\in\nu_{n}(t_0)}(\Delta_a{X}_{i,n})^2}{\sum_{i\in\nu_{n}(t_0)}(\Delta_a{X}_{i,n})^2}\\
&&= C v(n)^{H(t_0)}|\log(v(n))|^{1/2},
\end{eqnarray*}
where $C=C_5 (\partial_y\Phi(t_0,X(t_0))\big)^{-2}$ is a strictly positive random variable, due to the fact that $\partial_y\Phi(x,y)$ does not vanish almost surely over $\mathbb R_+\times\mathbb R\backslash\{0\}$ and $X(t_0)$, for $t_0\in(0,1)$, is a non degenerate Gaussian random variable. Lemma \ref{lemmevn3} is thus proved. $\square$

\begin{lemma}
\label{lemmeV13}
Assume that $v$ satisfies condition $(i)$ in Theorem~\ref{th:V1}. Then,
\begin{equation}
\label{diffv23}
\frac{V_{n}(t_0)}{V_{n,2}(t_0)}=1+ \mathcal O_{a.s.}\left(\bigg(\sum_{l=0}^4v(n)^ln^{(l-2)H(t_0)}|\log n|^{2-l/2}\bigg)^{1/2}\right).
\end{equation}
\end{lemma}
\textbf{Proof of Lemma \ref{lemmeV13}.} In view of (\ref{eq1:th:Vch5}), (\ref{incYbarantch5}) and (\ref{defYbarantch5}), $V_{n}(t_0)$ can be expressed as
\begin{equation}
\label{vn2pre}
V_{n}(t_0)=\sum_{i\in\nu_{n}(t_0)}\left(\sum_{k=0}^pa_k\Phi\Big(\frac{i+k}{n},X\big(\frac{i+k}{n}\big)\Big)\right)^2.
\end{equation}
For $t$ being in the neighborhood of $t_0$, the second order Taylor expansion of $\Phi(t,X(t))$ on $(t_0,X(i/n))$, the inequalities (\ref{fpol:antch5}) and (\ref{diffX}) yield that, for $t\in(0,1)$ satisfying $|t-t_0|=\mathcal O(v(n))$ and $|t-i/n|=\mathcal O(n^{-1})$, we have
\begin{eqnarray}
\label{fTaylor}
&&\Phi(t,X(t))=\Phi\Big(t_0,X\big(\frac{i}{n}\big)\Big)+\partial_x\Phi\Big(t_0,X\big(\frac{i}{n}\big)\Big)(t-t_0)\nonumber\\
&&~~~~~~~~+\partial_y\Phi\Big(t_0,X\big(\frac{i}{n}\big)\Big)\Big(X(t)-X\big(\frac{i}{n}\big)\Big)+e_{i,n}(t),
\end{eqnarray}
where, by (\ref{diffX}) and (\ref{diffHt}) we know that the remaining term $e_{i,n}(t)$ (depending on $i,n,t$) in the above equation satisfies
\begin{eqnarray}
\label{ein}
e_{i,n}(t)&=&\mathcal O_{a.s.}\left((X(t)-X\big(\frac{i}{n}\big))^2+\Big(X(t)-X\big(\frac{i}{n}\big)\Big)(t-t_0)+(t-t_0)^2\right)\nonumber\\
&=&\mathcal O_{a.s.}\left(n^{-2H(t_0)}|\log n|+v(n)n^{-H(t_0)}|\log n|^{1/2}+v(n)^2\right)\nonumber\\
&=&\mathcal O_{a.s.}\left(\sum_{l=0}^2v(n)^ln^{(l-2)H(t_0)}|\log n|^{1-l/2}\right).
\end{eqnarray}
Taking $t=(i+k)/n$ for $i\in\nu_n(t_0)$ and $k\in\{0,\ldots,p\}$ in (\ref{fTaylor}), and taking $e_{i,n}=\sum_{k=0}^pa_ke_{i,n}((i+k)/n)$ for short, in view of (\ref{vn2pre}) and  (\ref{moment:a:}), we obtain that
\begin{eqnarray}
\label{vn2pre1}
&&V_{n}(t_0)=\sum_{i\in\nu_{n}(t_0)}\left(\sum_{k=0}^pa_k\Big(\Phi\Big(t_0,X\big(\frac{i}{n}\big)\Big)+\partial_x\Phi\Big(t_0,X\big(\frac{i}{n}\big)\Big)\Big(\frac{i+k}{n}-t_0\Big)\right.\nonumber\\
&&\left.~~~~~~~~~~~~~~~+\partial_y\Phi\Big(t_0,X\big(\frac{i}{n}\big)\Big)\Big(X\big(\frac{i+k}{n}\big)-X\big(\frac{i}{n}\big)\Big)+e_{i,n}\big(\frac{i+k}{n}\big)\Big)\right)^2\nonumber\\
&&=\sum_{i\in\nu_{n}(t_0)}\Big(\partial_y\Phi\Big(t_0,X\big(\frac{i}{n}\big)\Big)\Delta_a{X}_{i,n}+e_{i,n}\Big)^2\nonumber\\
&&=V_{n,2}(t_0)+2\sum_{i\in\nu_{n}(t_0)}\partial_y\Phi\Big(t_0,X\big(\frac{i}{n}\big)\Big)\big(\Delta_a{X}_{i,n}\big)e_{i,n}+\sum_{i\in\nu_{n}(t_0)}e_{i,n}^2.
\end{eqnarray}
Using Cauchy-Schwarz inequality and (\ref{eq2:lemmevn3antch5}), we get
\begin{eqnarray}
\label{sum:fe}
&&\left|\sum_{i\in\nu_{n}(t_0)}\partial_y\Phi\Big(t_0,X\big(\frac{i}{n}\big)\Big)\Delta_a{X}_{i,n}e_{i,n}\right|\nonumber\\
&&\leq
\left(\sum_{i\in\nu_{n}(t_0)}(\partial_y\Phi\Big(t_0,X\big(\frac{i}{n}\big)\Big)\Delta_a{X}_{i,n})^2\right)^{1/2}\bigg(\sum_{i\in\nu_{n}(t_0)}e_{i,n}^2\bigg)^{1/2}\nonumber\\
&&=\big(V_{n,2}(t_0)\big)^{1/2}\bigg(\sum_{i\in\nu_{n}(t_0)}e_{i,n}^2\bigg)^{1/2}.
\end{eqnarray}
Putting together (\ref{vn2pre1}), (\ref{sum:fe}), (\ref{ein}), (\ref{eq1:lemmevn3antch5}), the fact that $(\partial_y\Phi(t_0,X(t_0)))^{-2}$ is an almost surely finite random variable, (\ref{eq4:ant-ch5}), (\ref{eq5:ant-ch5}), Proposition~\ref{lemmeVN2} and Lemma~\ref{lemmevn3}, we obtain that, for all $n\ge p+1$,
\begin{eqnarray}
\label{diffv23pre}
&&\frac{|V_{n}(t_0)-V_{n,2}(t_0)|}{V_{n,2}(t_0)}\le 2\frac{\big(\sum_{i\in\nu_{n}(t_0)}e_{i,n}^2\big)^{1/2}}{\big (V_{n,2}(t_0)\big)^{1/2}}+\frac{\sum_{i\in\nu_{n}(t_0)}e_{i,n}^2}{V_{n,2}(t_0)}\nonumber\\
&&=\frac{2\big(\sum_{i\in\nu_{n}(t_0)}e_{i,n}^2\big)^{1/2}}{\big((\partial_y\Phi(t_0,X(t_0)))^2v(n)n^{1-2H(t_0)}\big)^{1/2}} \frac{\big((\partial_y\Phi(t_0,X(t_0)))^2v(n)n^{1-2H(t_0)}\big)^{1/2}}{\big (V_{n,1}(t_0)\big)^{1/2}}\nonumber\\
&&\hspace{2cm}\times \Big(\frac{V_{n,1}(t_0)}{V_{n,2}(t_0)}\Big)^{1/2}+\frac{\sum_{i\in\nu_{n}} e_{i,n}^2}{(\partial_y\Phi(t_0,X(t_0)))^2v(n)n^{1-2H(t_0)}}\nonumber\\
&&\hspace{3cm}\times \frac{(\partial_y\Phi(t_0,X(t_0)))^2v(n)n^{1-2H(t_0)}}{V_{n,1}(t_0)}\times\frac{V_{n,1}(t_0)}{V_{n,2}(t_0)}\nonumber\\
&& = \mathcal O_{a.s.}\left(\bigg(\sum_{l=0}^4v(n)^ln^{(l-2)H(t_0)}|\log n|^{2-l/2}\bigg)^{1/2}\right).~~ \square
\end{eqnarray}
The following remark is a straightforward consequence of Lemma \ref{diffv23}, Lemma~\ref{lemmevn3} and Proposition~\ref{lemmeVN2}
\begin{remark}
\label{rem2:antch5}
Assume that $v$ satisfies the condition $(i)$ in Theorem~\ref{th:V1}, then we have,
\begin{eqnarray}
\label{eq1rem2:antch5}
&&\frac{V_{n}(t_0)}{2\widetilde{C}_a(H(t_0))(\partial_y\Phi(t_0,X(t_0)))^2 v(n)n^{1-2H(t_0)}}-1\nonumber\\
&&=\mathcal O_{a.s.}\left(\left|\frac{V_n(t_0)}{V_{n,2}(t_0)}-1\right|+\left|\frac{V_{n,2}(t_0)}{V_{n,1}(t_0)}-1\right|\right)\nonumber\\
&&~~~~+\mathcal O_{\mathbb P}\left(\left|\frac{V_{n,1}(t_0)}{2\widetilde{C}_a(H(t_0))(\partial_y\Phi(t_0,X(t_0)))^2 v(n)n^{1-2H(t_0)}}-1\right|\right)\nonumber\\
&&=\mathcal O_{a.s.}\left(\bigg(\sum_{l=0}^4v(n)^ln^{(l-2)H(t_0)}|\log n|^{2-l/2}\bigg)^{1/2}+v(n)^{H(t_0)}|\log(v(n))|^{1/2}\right)\nonumber\\
&&~~~~+\mathcal O_{\mathbb P}\left(v(n)\log n+v(n)^{-1}n^{-1}\right).
\end{eqnarray}
Moreover, if the condition $(ii)$ in Theorem~\ref{th:V1} is satisfied, then by using  (\ref{marcov1}) in Section \ref{Appendix} we get existence of a constant $c>0$ such that
$$
\sum_{n=p+1}^{\infty}\mathbb P\left(\bigg|\frac{\sum_{i\in\nu_{n}(t_0)}(\Delta_a{X}_{i,n})^2}{\sum_{i\in\nu_{n}(t_0)}Var(\Delta_a{X}_{i,n})}-1\bigg|>\eta\right)\le c\eta^{-4},~\mbox{for all $\eta>0$}.
$$
It then follows from Borel-Cantelli lemma that
$$
\frac{\sum_{i\in\nu_{n}(t_0)}(\Delta_a{X}_{i,n})^2}{\sum_{i\in\nu_{n}(t_0)}Var(\Delta_a{X}_{i,n})}\xrightarrow[n\to\infty]{a.s.}1,
$$
and thus
\begin{equation}
\label{eq1rem2:antch10}
\frac{V_{n}(t_0)}{2\widetilde{C}_a(H(t_0))(\partial_y\Phi(t_0,X(t_0)))^2 v(n)n^{1-2H(t_0)}}\xrightarrow[n\rightarrow\infty]{a.s.}1.
\end{equation}
\end{remark}
Now we are ready prove Theorem~\ref{th:V1}.\\
\textbf{ Proof of Theorem \ref{th:V1}.} Let $g:(0,+\infty)\rightarrow{\mathbb R}$, be the continuous function defined for
all $x\in (0,+\infty)$, as:
\begin{equation}
\label{eq9:th:Vch5}
g(x)=\frac{1}{2}\big(1+\log_2(x)\big).
\end{equation}
Observe that, in view of (\ref{eq2:th:Vch5}),  for all $n\ge p+1$, we have
\begin{equation}
\label{eq11:th:Vch5}
\widehat{H}_{n,t_0}=g\left(\frac{v(2n)V_{n}(t_0)}{v(n)V_{2n}(t_0)}\right).
\end{equation}
On the other hand observe that (\ref{eq1rem2:antch5}) implies that,
\begin{equation}
\label{eq12:th:Vch5}
\widehat{H}_{n,t_0}-H(t_0)=g\left(\frac{v(2n)V_{n}(t_0)}{v(n)V_{2n}(t_0)}\right)-g\left(2^{2H(t_0)-1)}\right).
\end{equation}
Thus combining (\ref{eq11:th:Vch5}), (\ref{eq12:th:Vch5}), (\ref{eq1rem2:antch5}) and the fact that $g$ is continuously differentiable on $2^{2H(t_0)-1}$, we obtain that,
\begin{eqnarray*}
&&\widehat{H}_{n,t_0}-H(t_0)\nonumber\\
&&=\mathcal O_{a.s.}\left(\bigg(\sum_{l=0}^4v(n)^ln^{(l-2)H(t_0)}|\log n|^{2-l/2}\bigg)^{1/2}+v(n)^{H(t_0)}|\log(v(n))|^{1/2}\right)\nonumber\\
&&~~~~+\mathcal O_{\mathbb P}\left(v(n)\log n+v(n)^{-1}n^{-1}\right).
\end{eqnarray*}
This proves the first statement \textbf{(1)} in Theorem \ref{th:V1}. 
Next if $v(n)$ further satisfies the condition $(ii)$, then similarly starting from (\ref{eq1rem2:antch10}) we get
$$
\widehat{H}_{n,t_0}\xrightarrow[n\rightarrow\infty]{a.s.}H(t_0).
$$
the statement \textbf{(2)} in Theorem \ref{th:V1} is proved. $\square$

\section{Appendix}
\label{Appendix}
\subsection{Proof of Proposition \ref{maj:cov}}
\subsubsection{Proof of (\ref{kneqk'})} 
 By using the isometry property of mBm's harmonizable presentation, the fact that $\sigma$ and $X$ are independent leads to that  the $\Delta_aY_{k,n}$'s are zero-mean random variables, we then get
\begin{eqnarray}
\label{cov1}
&&Cov(\Delta_aY_{k,n} ,\Delta_aY_{k',n})\nonumber\\
&&=\sum_{j=0}^p\sum_{j'=0}^pa_ja_{j'}\mathbb E\Big(\sigma\big(\frac{j+k}{n}\big)\sigma\big(\frac{j'+k'}{n}\big)\Big)\mathbb E\Big( X\big(\frac{j+k}{n}\big)X\big(\frac{j'+k'}{n}\big)\Big)\nonumber\\
&&=\sum_{j=0}^p\sum_{j'=0}^pa_ja_{j'}\theta\big(\frac{j+k}{n},\frac{j'+k'}{n}\big)\int_{\mathbb{R}}\frac{\big(
e^{i(\frac{j+k}{n}) u}-1\big) \big( e^{-i( \frac{j'+k'}{n})
u}-1\big) }{|u| ^{H( \frac{j+k}{n}) +H(
\frac{j'+k'}{n}) +1}}{\,\mathrm{d}} u{\,\mathrm{d}} t{\,\mathrm{d}} s.\nonumber\\
\end{eqnarray}
Since the PHE of $X$ in the neighborhood of $k/n,~k'/n$ behave
locally asymptotically like those of fBms with Hurst parameters $H(k/n),H(k'/n)$ respectively, we can thus consider a Taylor expansion
of $f$ around $(k/n,k'/n)$. Fix $u\in\mathbb R$ and define $f(x,y) =\theta(x,y)|u| ^{-H(x)-H(y) -%
1}$, since $f$ belongs to $C^2([0,1]^2)$, we take the second order Taylor expansion of the bivariate function $f$ on $(
k/n,k'/n)$: there exist $\xi_{j,k,n} \in ( k/n,(j+k)/n) $ and $%
\xi'_{j',k',n}\in ( k'/n,(j'+k')/n) $ such that
\begin{eqnarray}
\label{A}
&&f\big( \frac{j+k}{n},\frac{j'+k'}{n}\big)=f\big( \frac{k}{n},\frac{k'}{n}\big)+\partial_xf\big( \frac{k}{n},\frac{k'}{n}\big)\big(\frac{j}{n}\big)+\partial_yf\big( \frac{k}{n},\frac{k'}{n}\big)\big(\frac{j'}{n}\big)\nonumber\\
 &&\hspace{1cm}+\partial_{xy}f\big( \frac{k}{n},\frac{k'}{n}\big)\big(\frac{jj'}{n^2}\big)+\frac{1}{2}\partial_{x}^2f\big( \xi_{j,k,n},\frac{k'}{n}\big)\big(\frac{j}{n}\big)^2+\frac{1}{2}\partial_{y}^2f\big( \frac{k}{n},\xi'_{j',k',n}\big)\big(\frac{j'}{n}\big)^2\nonumber\\
 &&=\sum_{l,l'\in\{0,1,2\},l+l'\le 2}A_{l,l'}(u,j,j',k,k',n),
\end{eqnarray}%
where we denote, for $u\neq 0$, $x,y\ge0$,
$$A_{0,0}(u,j,j',k,k',n)=f\big( \frac{k}{n},\frac{k'}{n}\big)=\theta\big( \frac{k}{n},\frac{k'}{n}\big)|u| ^{-H(k/n)-H(k'/n) -1};
$$
\begin{eqnarray*}
&&A_{0,1}(u,j,j',k,k',n)\\
&&=|u|^{-H(k/n)-H(k'/n)-1}\Big(\partial_y\theta\big(\frac{k}{n},\frac{k'}{n}\big)-\theta\big(\frac{k}{n},\frac{k'}{n}\big)H'\big(\frac{k'}{n}\big)
\log|u|\Big)\big(\frac{j'}{n}\big);\\
&&A_{1,0}(u,j,j',k,k',n)\\
&& =|u|^{-H(k/n)-H(k'/n)-1}\Big(\partial_x\theta\big(\frac{k}{n},\frac{k'}{n}\big)-\theta\big(\frac{k}{n},\frac{k'}{n}\big)H'\big(\frac{k}{n}\big)
\log|u|\Big)\big(\frac{j}{n}\big);
\end{eqnarray*}
\begin{eqnarray*}
&&A_{1,1}(u,j,j',k,k',n)\\
&& =|u|^{-H(k/n)-H(k'/n)-1}\Big(\partial_{xy}\theta\big(\frac{k}{n},\frac{k'}{n}\big)-\partial_x\theta\big(\frac{k}{n},\frac{k'}{n}\big)H'\big(\frac{k'}{n}\big)\log|u|\\
&&\quad\quad-\partial_y\theta\big(\frac{k}{n},\frac{k}{n}\big)H'\big(\frac{k'}{n}\big)\log|u|
+\theta\big(\frac{k}{n},\frac{k'}{n}\big)H'\big(\frac{k}{n}\big)H'\big(\frac{k'}{n}\big)\big(\log|u|\big)^2\big(\frac{jj'}{n^2}\big);\\
\end{eqnarray*}
and
\begin{eqnarray*}
&&A_{0,2}(u,j,j',k,k',n)=\frac{1}{2}|u|^{-H(k/n)-H(k'/n)-1}\big(\frac{j'}{n}\big)^2\\
&&~\times\bigg(\Big(\partial_y^2\theta\big(\frac{k}{n},\xi'_{j',k',n}\big)
-\Big(2\partial_y\theta\big(\frac{k}{n},\frac{k'}{n}\big)H'\big(\frac{k'}{n}\big)+\theta\big(\frac{k}{n},\frac{k'}{n}\big)H''\big(\xi'_{j',k',n}\big)\Big)\log|u|\\
&&\hspace{3cm}+\theta\big(\frac{k}{n},\frac{k'}{n}\big)\Big(H'\big(\frac{k'}{n}\big)\Big)^2\big(\log|u|\big)^2\bigg);\\
&&A_{2,0}(u,j,j',k,k',n)=\frac{1}{2}|u|^{-H(k/n)-H(k'/n)-1}\big(\frac{j}{n}\big)^2\\
&&\quad\quad\times\bigg(\Big(\partial_x^2\theta\big(\xi_{j,k,n},\frac{k'}{n}\big)
-\Big(2\partial_x\theta\big(\frac{k}{n},\frac{k'}{n}\big)H'\big(\frac{k}{n}\big)+\theta\big(\frac{k}{n},\frac{k'}{n}\big)H''\big(\xi_{j,k,n}\big)\Big)\log|u|\\
&&\hspace{3cm}+\theta\big(\frac{k}{n},\frac{k'}{n}\big)\Big(H'\big(\frac{k}{n}\big)\Big)^2\big(\log|u|\big)^2\bigg).
\end{eqnarray*}
Thus we rewrite (\ref{cov1}) as
\begin{equation}
\label{covYY}
Cov\big(\Delta_aY_{k,n},\Delta_aY_{k',n}\big)=\sum\limits_{{}^{l,l'\in \{ 0,1,2\},}_{l+l'\le2} }\mathcal{I}%
_{l,l'}( k,k',n),
\end{equation}
where
$$\mathcal{I}_{l,l'}( k,k^{\prime },n)=\sum_{j=0}^p\sum_{j'=0}^pa_ja_{j'}\int_{\mathbb{R}}\big( e^{i\left( \frac{j+k}{n}\right) u}-1\big) \big(
e^{-i\left( \frac{j'+k'}{n}\right) u}-1\big) A_{l,l'}\ud u.
$$
We note here $A_{l,l'}$'s are notations in short for $A_{l,l'}(u,j,j',k,k')$ in (\ref{A}).
By using (\ref{cov1}), it suffices to make an identification of all the terms $\mathcal{I}_{l,l^{\prime
}}\left( k,k^{\prime },n\right)$'s in order to estimate the covariance structure of the wavelet coefficients. We consider different cases according to the values of $(l,l')$. The key to these identifications is to observe the following:\begin{itemize}
                        \item First, observe that for $x,y>0,\ x\neq y,\ \alpha >0$, $p\in\mathbb N$, we have
\begin{eqnarray}
\label{general}
&&\int_{\mathbb{R}}\frac{( e^{ixu}-1)(
e^{-iyu}-1) }{\left\vert u\right\vert ^{\alpha+1 }}(\log|u|)^p {\,\mathrm{d}} u=\frac{1}{2}\sum_{l=0}^p(-1)^lC_{p-l}(\alpha)\nonumber\\
&&\hspace{1cm}\times\big( \left\vert x\right\vert ^{\alpha}(\log \left\vert
x\right\vert)^l +\left\vert y\right\vert ^{\alpha}(\log \left\vert
y\right\vert)^l -\left\vert x-y\right\vert ^{\alpha }(\log \left\vert
x-y\right\vert)^l \big), \nonumber \\
\end{eqnarray}
where  for $l\in\{0,\ldots,p\}$, $C_{l}( \alpha ) ={p \choose l}\int_{\mathbb{R}}\frac{| e^{iu}-1| ^{2}}{\left\vert u\right\vert ^{\alpha+1 }} (\log \left\vert u\right\vert)^l {\,\mathrm{d}} u$, with ${p\choose l}=\frac{p!}{l!(p-l)!}$ being the binomial coefficient.
                        \item Secondly, for $k\neq k'$, $Q\ge1$, $l,l'\in\{0,\ldots,Q\}$, $p\in\mathbb N$ and $\alpha>0$, a $2Q-l-l'$ order Taylor expansion of $q_{\alpha,p}(x)=(1+x)^{\alpha}(\log|1+x|)^p$ on $x=\frac{j-j'}{k-k'}$ yields:
\begin{eqnarray}
\label{computeI}
&&%
\sum_{j=0}^p\sum_{j'=0}^pa_ja_{j'}j^l{j'}^{l'}
\Big|\frac{j+k-j'-k'}{n}\Big| ^{\alpha }\Big(\log\big|\frac{j+k-j'-k'}{n}\big|\Big)^p\nonumber\\
&&=\Big|\frac{k-k'}{n}\Big| ^{\alpha }\sum_{v=0}^p{v \choose p}\Big(\log\big|\frac{k-k'}{n}\big|\Big)^{p-v}\sum_{j=0}^p\sum_{j=0}^pj^l{j'}^{l'}
 \Big( 1+\frac{j-j'}{k-k'}\Big) ^{\alpha }\nonumber\\
&&\hspace{3cm}\times\Big(\log\big| 1+\frac{j-j'}{k-k'}\big|\Big)^{v}\nonumber\\
&&=\Big|\frac{k-k'}{n}\Big| ^{\alpha}\sum_{v=0}^p{p\choose v}\Big(\log\big|\frac{k-k'}{n}\big|\Big)^{p-v}\sum_{j=0}^p\sum_{j'=0}^pj^l{j'}^{l'}\nonumber\\
&&\hspace{3cm}\times \theta_{\alpha,2Q-l-l',v}\big(\frac{j-j'}{k-k'}\big)\Big(\frac{j-j'}{k-k'}\Big) ^{2Q-l-l'}\nonumber\\
&&=\frac{n^{-\alpha}}{|k-k'|^{2Q-l-l'-\alpha}}\bigg(\sum_{v=0}^p{p\choose v}A_{\alpha,2Q-l-l',v}\Big(\log\big|\frac{k-k'}{n}\big|\Big)^{p-v}\bigg), \nonumber\\
\end{eqnarray}
where the integral remainder $\theta_{\alpha,2Q-l-l',v}$ of $q(\cdot)$'s $(2Q-l-l')$-th order Taylor expansion (see e.g. \cite{Apostol1967}) is given as:
 \begin{itemize}
 \item for $2Q-l-l'=0$,
\begin{equation}
\label{theta'}
\theta_{\alpha,2Q-l-l',v}\big(\frac{j-j'}{k-k'}\big)=q_{\alpha,v}\big(\frac{j-j'}{k-k'}\big);
\end{equation}
\item for $2Q-l-l'\ge1$,
\begin{eqnarray}
\label{theta}
&&\theta_{\alpha,2Q-l-l',v}\Big(\frac{j-j'}{k-k'}\Big) = \frac{1 }{%
(2Q-l-l'-1) !}\nonumber\\
&&\times\int_0^1\!\!(1-\eta)^{2Q-l-l'-1}q_{\alpha,v}^{(2Q-l-l')}\Big(\eta \frac{j-j'}{k-k'}+(1-\eta)\Big){\,\mathrm{d}}\eta; \nonumber\\
\end{eqnarray}
\end{itemize}
and the term $A_{\alpha,2Q-l-l',v}$ is defined to be
\begin{equation}
\label{Aab}
A_{\alpha,2Q-l-l',v}=\sum_{j=0}^p\sum_{j'=0}^pa_ja_{j'}\theta_{\alpha,2Q-l-l',v}\big(\frac{j-j'}{k-k'}\big)\big(j-j'\big)^{2Q-l-l'}.
\end{equation}
Here we note that for $r\in\mathbb N$ and $|x|< 1$,
$$
q_{\alpha,p}^{(r)}(x)=(1+x)^{\alpha-r}\!\!\!\!\sum_{u,v\in\mathbb N,u+v=r}\!\!\!\!(\alpha-1)\ldots(\alpha-u)p\ldots(p-v)(\log|1+x|)^{p-v}.
$$
                      \end{itemize}
\begin{description}
\item[Case (i)] $l=l^{\prime }=0 $.

In this case we have
$$\mathcal{I}_{0,0}\left( k,k^{\prime
},n\right)=\theta\big(\frac{k}{n},\frac{k'}{n}\big)\int_{\mathbb{R}}\frac{( e^{i\left( \frac{j+k}{n}\right) u}-1)
( e^{-i\left( \frac{j'+k'}{n}\right) u}-1) }{\left\vert
u\right\vert ^{H\left( k/n\right) +H\left( k'/n\right) +1}}{\,\mathrm{d}} u.
$$
Let $p=0$, $x=(j+k)/n,\ y=(j'+k')/n$ and $\alpha =H(k/n)+H(k'/n)$ in (\ref{general}). It follows
\begin{eqnarray*}
&&\mathcal{I}_{0,0}\left( k,k^{\prime },n\right)=-\frac{C_0(H(k/n)+H(k'/n))}{2}n^{-H(k/n)-H(k'/n)}\nonumber\\
&&\hspace{2cm}\times\theta\big(\frac{k}{n},\frac{k'}{n}\big)\sum_{j=0}^p\sum_{j'=0}^na_ja_{j'} \left\vert j+k-j'-k^{\prime }\right\vert ^{H\left( k/n\right) +H(
k^{\prime }/n) }.
\end{eqnarray*}%
Then by the assumption $k\neq k'$, we can thus take $l=l'=0$ in (\ref{computeI}) to obtain
\begin{eqnarray*}
&&\mathcal{I}_{0,0}\left( k,k^{\prime },n\right)=-\frac{C_0(H(k/n)+H(k'/n))}{2}\nonumber\\
&&~~~~\times n^{-H(k/n)-H(k'/n)}A_{H(k/n)+H(k'/n),2Q,0}\frac{\theta(k/n,k'/n)}{%
\left\vert k-k^{\prime }\right\vert ^{2Q-H(k/n)-H(k'/n)}}.\nonumber\\
\end{eqnarray*}
We finally obtain
\begin{equation}
\label{computeI00}
\mathcal{I}_{0,0}\left( k,k^{\prime },n\right) =C\Big(H\big(\frac{k}{n}\big)+H\big(\frac{k'}{n}\big),Q\Big)\frac{\theta(k/n,k'/n)n^{-H(k/n)-H(k'/n)}}{\left\vert k-k^{\prime }\right\vert ^{2Q-H(k/n)-H(k'/n) }},
\end{equation}%
where the coefficient $C(H(k/n)+H(k'/n),Q)$ is given by
\begin{equation}
\label{def:C}
C\Big(H\big(\frac{k}{n}\big)+H\big(\frac{k'}{n}\big),Q\Big)=-\frac{C_0(H(k/n)+H(k'/n))}{2}A_{H(k/n)+H(k'/n),2Q,0}.
\end{equation}
\item[Case (ii)] $(l,l')\in\{(1,0),(0,1)\}.$
Since $\theta\in C^2([0,1]^2)$ and $H\in C^{2}([0,1])$, then by definition of $\mathcal{I}_{1,0}( k,k^{\prime },n)$ and the triangle inequality, $\mathcal{I}_{1,0}( k,k^{\prime },n)$ can be expressed as
$$\sum_{v=0}^1\mathcal O\Big(\sum_{j=0}^p\sum_{j'=0}^p \big(\frac{j}{n}\big)\int_{\mathbb{R}}\frac{(
e^{i( \frac{j+k}{n}) u}-1)( e^{-i( \frac{j'+k'}{n}) u}-1) }{|u| ^{H(k/n)
+H(k'/n) +1}}(\log |u|)^v {\,\mathrm{d}} u\Big).
$$
 In (\ref{general}), we let $p=0$ and $p=1$ (respectively corresponding to $v=0$ and $v=1$ of the above expression), $x=(j+k)/n, y=(j'+k')/n$ and $\alpha =H(k/n)+H( k'/n)$. And take $(l,l')=(1,0)$ in (\ref{computeI}), then
\begin{equation*}
\label{computeI10}
\mathcal{I}_{1,0}\left( k,k^{\prime },n\right)=\mathcal O\Big(\frac{n^{-H(k/n)-H(k'/n)-1}(1+|\log|(k-k')/n||)}{|k-k'|^{2Q-H(k/n)-H(k'/n)-1}}\Big).
\end{equation*}
Similarly we can get at the meanwhile,
\begin{equation*}
\mathcal{I}_{0,1}\left( k,k^{\prime },n\right)=\mathcal O\Big(\frac{n^{-H(k/n)-H(k'/n)-1}(1+|\log|(k-k')/n||)}{|k-k'|^{2Q-H(k/n)-H(k'/n)-1}}\Big).
\end{equation*}
We conclude that 
\begin{equation}
\label{computeI01}
\sum_{{}^{l,l'\in\{0,1,2\}}_{l+l'=1}}\mathcal{I}_{l,l'}\left( k,k',n\right)=\mathcal O\Big(\frac{n^{-H(k/n)-H(k'/n)}(1+|\log|(k-k')/n||)}{|k-k'|^{2Q-H(k/n)-H(k'/n)-1}}\Big).
\end{equation}
\item[Case (iii)] $l=l'=1.$

Observe that $\mathcal{I}_{1,1}\left( k,k^{\prime },n\right)$ can be expressed as
$$\sum_{v=0}^2\mathcal O\Big(\sum_{j=0}^p\sum_{j'=0}^pa_ja_{j'}\big(\frac{jj'}{n^2}\big)\int_{\mathbb{R}} \frac{\big(
e^{i( \frac{j+k}{n}) u}-1\big) \big( e^{-i( \frac{j'+k'}{n}) u}-1\big) }{|u| ^{H(k/n)
+H(k'/n) +1}}(\log |u|)^v {\,\mathrm{d}} u\Big).
$$

 Let $p=0,1,2$ respectively, $x=(j+k)/n,\ y=(j'+k')/n$ and $\alpha =H\left( k/n \right)
+H\left( k'/n\right)$ in (\ref{general}) and $(l,l')=(1,1)$ in (\ref{computeI}), we get
\begin{equation}
\label{computeI11}
\mathcal{I}_{1,1}\left( k,k^{\prime },n\right)=\sum_{v=0}^2\mathcal O\Big(\frac{n^{-H(k/n)-H(k'/n)-2}}{|k-k'|^{2Q-H(k/n)-H(k'/n)-2}}\Big|\log\big|\frac{k-k'}{n}\big|\Big|^v\Big).
\end{equation}
\item[Case (iv)] $(l,l')\in\{(2,0),(0,2)\}.$\\
Still by using the facts that $\theta\in C^2([0,1]^2)$ and $H\in C^2([0,1])$, $\mathcal{I}_{2,0}\left( k,k^{\prime },n\right)$ can be expressed as
$$
\sum_{p=0}^2\!\mathcal O\Big(n^{-2}\sum_{j=0}^p\Big|\sum_{j'=0}^pa_{j'}\int_{\mathbb{R}} \frac{(
e^{i( \frac{j+k}{n}) u}-1) ( e^{-i( \frac{j'+k'}{n}) u}-1) }{|u| ^{H(\xi_{j,k,n})
+H(k/n') +1}}(\log|u|)^p {\,\mathrm{d}} u\Big|\Big).
$$
 Let $p=0,1,2$ respectively, and let $x=(j+k)/n, y=(j'+k')/n$ and $\alpha =H\left( \xi_{j,k,n} \right)
+H\left( k^{\prime }/n\right)$ in (\ref{general}), we get
\begin{eqnarray*}
&&\mathcal{I}_{2,0}\left( k,k^{\prime },n\right)=\sum_{p=0}^2\mathcal O\Big(n^{-2}\int_{0}^{1}\Big|\sum_{j=0}^pa_j\sum_{v=0}^p(-1)^vC_{p-v}(\alpha)\\
&&\times\bigg( \Big| \frac{j'+k'}{n}\Big| ^{\alpha}\Big(\log \big|
\frac{j'+k'}{n}\big|\Big)^v\\
&&\hspace{3cm}-\Big|\frac{j-j'+k-k'}{n}\Big| ^{\alpha }\Big(\log \big|\frac{j-j'+k-k'}{n}\big|\Big)^v\Big) {\,\mathrm{d}} u\bigg).
\end{eqnarray*}
Then similarly to (\ref{computeI}), using a $Q$ order Taylor expansion of $q_{\alpha,p}(\cdot)$ respectively on $j'/k'$ and on $(j-j')/(k-k')$, and also use the fact that for $x>0$, $x^{H(\xi_{j,k,n})}\sim x^{H(k/n)}$ (since $\xi_{j,k,n}\in(k/n,(j+k)/n)$), we obtain
\begin{eqnarray}
\label{computeI20}
&&\mathcal{I}_{2,0}\left( k,k^{\prime },n\right)=\mathcal O\left(\frac{n^{-H(k/n)-H(k'/n)-2}}{|k'|^{Q-H(k/n)-H(k'/n)}}\left(1+\Big|\log\big|\frac{k'}{n}\big|\Big|+\Big|\log\big|\frac{k'}{n}\big|\Big|^2\right)\right)\nonumber\\
&&+\mathcal O\left(\frac{n^{-H(k/n)-H(k'/n)-2}}{|k-k'|^{Q-H(k/n)-H(k'/n)}}\left(1+\Big|\log\big|\frac{k-k'}{n}\big|\Big|+\Big|\log\big|\frac{k-k'}{n}\big|\Big|^2\right)\right).\nonumber\\
\end{eqnarray}
By using similar principle,
\begin{eqnarray}
\label{computeI02}
&&\mathcal{I}_{0,2}\left( k,k^{\prime },n\right)=\mathcal O\left(\frac{n^{-H(k/n)-H(k'/n)-2}}{|k|^{Q-H(k/n)-H(k'/n)}}\left(1+\Big|\log\big|\frac{k}{n}\big|\Big|+\Big|\log\big|\frac{k}{n}\big|\Big|^2\right)\right)\nonumber\\
&&+\mathcal O\left(\frac{n^{-H(k/n)-H(k'/n)-2}}{|k-k'|^{Q-H(k/n)-H(k'/n)}}\left(1+\Big|\log\big|\frac{k-k'}{n}\big|\Big|+\Big|\log\big|\frac{k-k'}{n}\big|\Big|^2\right)\right).\nonumber\\
\end{eqnarray}
\end{description}
Now denote by
\begin{equation}
\label{Tkk'}
R(n,k,k',Q)=n^{H(k/n)+H(k'/n)}\sum_{{}^{l,l'\in\{0,1,2\},}_{1\le l+l'\le 2}}\mathcal{I}_{l,l'}(k,k',n).
\end{equation}
It remains to show that for $Q\ge2$,
$$
\sum_{0\le k,k'\le n}\big(R(n,k,k',Q)\big)^2=\mathcal O(n^{-1}(\log n)^2),~\mbox{as $n\to\infty$}.
$$
According to (\ref{Tkk'}), it suffices to prove for any $(l,l')\neq (0,0)$,
\begin{equation}
\label{boundI}
\sum_{0\le k,k'\le n}n^{2H(k/n)+2H(k'/n)}\big(\mathcal{I}_{l,l'}(k,k',n)\big)^2=\mathcal O(n^{-1}(\log n)^2).
\end{equation}
To  prove (\ref{boundI}) holds we only take $\mathcal I_{1,0}(k,k',n)$ as an example, since the computations for the other items are similar. Recall that
$$
\frac{\big(\mathcal I_{1,0}(k,k',n)\big)^2}{n^{-2H(k/n)-2H(k'/n)}}=\mathcal O\Big(\frac{n^{-2}\big(1+2|\log n|+|\log|k-k'||\big)^2}{|k-k'|^{4Q-2H(k/n)-2H(k'/n)-2}}\Big).
$$
    Remember that, the fact that $Q\ge2$ yields $4Q-2H(k/n)-2H(k'/n)-2>2$. This implies
\begin{eqnarray*}
&&\sum_{0\le k,k'\le n}n^{2H(k/n)+2H(k'/n)}\big(\mathcal I_{1,0}(k,k',n)\big)^2\\
&&=\mathcal O\Big(n^{-2}|\log n|^2\sum_{k=0}^{n}\sum_{|l|=1}^{\infty}\frac{(\log|l|)^2}{|l|^{4Q-4\sup_{t\in[0,1]}H(t)-2}}\Big)\\
&&=\mathcal O\big(n^{-1}|\log n|^2\big).
\end{eqnarray*}
The equation (\ref{kneqk'}) is proved. $\square$
\subsubsection{Proof of (\ref{k=k'})} 
When $k=k'$, we still have, as (\ref{covYY}), that
\begin{equation}
\label{covYY1}
Var\big(\Delta_aY_{k,n}\big)=\sum\limits_{l,l^{\prime }\in \{ 0,1,2\},l+l'\le2 }\mathcal{I}%
_{l,l'}( k,k,n).
\end{equation} 
Note that in this particular case $\mathcal I_{l,l'}(k,k,n)$ becomes

$$\mathcal{I}_{0,0}\left( k,k,n\right)=\theta\big(\frac{k}{n},\frac{k}{n}\big)\int_{\mathbb{R}}\frac{| e^{i\left( \frac{j+k}{n}\right) u}-1|^2
 }{\left\vert
u\right\vert ^{2H\left( k/n\right) +1}}\ud u.
$$
Let $p=0$, $x=y=(j+k)/n$ and $\alpha =2H(k/n)$ in (\ref{general}), we get
\begin{equation}
\label{prop1}
\mathcal{I}_{0,0}\left( k,k,n\right)=C_1\big(\frac{k}{n}\big)n^{-2H(k/n)},
\end{equation}
where the coefficient $C_1(k/n)$ is defined by:
\begin{equation}
\label{def:C1}
C_1\big(\frac{k}{n}\big)=-\frac{C_0(2H(k/n))}{2}n^{-2H(k/n)}\theta\big(\frac{k}{n},\frac{k}{n}\big)\sum_{j=0}^p\sum_{j'=0}^na_ja_{j'} \left\vert j-j'\right\vert ^{2H( k/n)}.
\end{equation}
The remaining items of $I_{l,l'}$'s are of higher order so it suffices to give proper upper bounds to them. By means of a Taylor expansion of $x\longmapsto x^\alpha (\log |x|^p)$ on $j-j'$,
similar discussion to the proof of  (\ref{kneqk'}) shows that, for $1\le l+l'\le 2$,
\begin{equation}
\label{prop11}
\mathcal{I}_{l,l'}\left( k,k,n\right)=\mathcal O\big(n^{-2H(k/n)-1}|\log n|^4\big).
\end{equation}
Hence (\ref{k=k'}) is proved by combining (\ref{prop1}) and (\ref{prop11}). $\square$

\subsection{Proof of Lemma \ref{iden:X}}
First by using the fact that
$$
Var\Big(\sum_{i\in\nu_{n}(t_0)}(\Delta_a{X}_{i,n})^2\Big)=\sum_{i,j\in\nu_{n}(t_0)}Cov\Big ((\Delta_a{X}_{i,n})^2,(\Delta_a{X}_{j,n})^2\Big ),
$$
and by using Lemma~\ref{lemmegauss} (in which we take $Z_1=\frac{\Delta_a{X}_{i,n}}{\sqrt{\Delta_a{X}_{i,n}}}$,
$Z_2=\frac{\Delta_a{X}_{j,n}}{\sqrt{\Delta_a{X}_{j,n}}}$ and $\tau=1$), we get
\begin{eqnarray}
\label{varsum:}
&&Var\Big(\sum_{i\in\nu_{n}(t_0)}(\Delta_a{X}_{i,n})^2\Big)\nonumber\\
&&=\sum_{i,j\in\nu_{n}(t_0)}\bigg(\mathbb E\Big(\Delta_a{X}_{i,n}\Delta_a{X}_{j,n}\Big)^2
-\mathbb E(\Delta_a{X}_{i,n})^2\mathbb E(\Delta_a{X}_{j,n})^2\bigg)\nonumber\\
&&=\sum_{i,j\in\nu_{n}(t_0)}Var(\Delta_a{X}_{i,n})Var(\Delta_a{X}_{j,n})\nonumber\\
&&\hspace{3cm}\times\bigg(\mathbb E\Big(\frac{\Delta_a{X}_{i,n}
\Delta_a{X}_{j,n}}{\sqrt{Var(\Delta_a{X}_{i,n})Var(\Delta_a{X}_{j,n})}}\Big)^2-1\bigg)\nonumber\\
&&=2\sum_{i,j\in\nu_{n}(t_0)}Var(\Delta_a{X}_{i,n})Var(\Delta_a{X}_{j,n})\bigg(Cov\bigg(\frac{\Delta_a{X}_{i,n}}{\sqrt{\Delta_a{X}_{i,n}}},
\frac{\Delta_a{X}_{j,n}}{\sqrt{\Delta_a{X}_{j,n}}}\bigg)\bigg)^2\nonumber\\
&&=2\sum_{i,j\in\nu_{n}(t_0)}\Big(Cov(\Delta_a{X}_{i,n},\Delta_a{X}_{j,n})\Big)^2,
\end{eqnarray}
Then it follows from (\ref{varsum:}) and Proposition~\ref{maj:cov} that there is a constant $c_1>0$ such that for all $n\ge p+1$,
\begin{eqnarray}
\label{varsum1}
Var\Big(\sum_{i\in\nu_{n}(t_0)}(\Delta_a{X}_{i,n})^2\Big)\le c_1\sum_{i,j\in\nu_{n}(t_0)}\frac{n^{-2(H(i/n)+H(j/n))}}{\big (1+|i-j|)^2}.
\end{eqnarray}
On the other hand, since $H\in C^2([0,1])$, there is a constant $c_2>0$ such that
for any $s,s'\in[0,1]$,
\begin{equation}
\label{Hc} |H(s)-H(s')|\leq c_2|s-s'|.
\end{equation}
Thus, it follows from (\ref{eq9:ant-ch5}) that for all $n\ge p+1$ and $i\in\nu_{n}(t_0)$,
\begin{equation}
\label{diffH}
n^{H(t_0)-H(i/n)}\leq  n^{|H(t_0)-H(i/n)|} \leq  n^{c_2|t_0-i/n|} \leq n^{c_2 v(n)}\le c_3,
\end{equation}
where $c_3 =\sup_{n\in\mathbb N,\, n\ge p+1} e^{c_2 v(n)\log n}$ is finite thanks to the condition $(ii)$ (see Theorem~\ref{th:V1}). Also (\ref{Hc}) and (\ref{eq9:ant-ch5}) imply
that
\begin{equation}
\label{diffHbis-ant}
n^{H(t_0)-H(i/n)}\ge n^{-|H(t_0)-H(i/n)|} \ge n^{-c_2 v(n)} \ge c_{3}',
\end{equation}
where $c_{3}'=\inf_{n\in\mathbb N,\,n\ge p+1} e^{-c_2 v(n)\log n}$ is strictly positive thanks to the condition $(ii)$ in Theorem~\ref{th:V1}.
It follows from (\ref{varsum1}), (\ref{diffH}), (\ref{eq4:ant-ch5}), (\ref{eq5:ant-ch5}) that for all $n\ge p+1$,
\begin{eqnarray}
\label{varsum2}
&&Var\Big(\sum_{i\in\nu_{n}(t_0)}(\Delta_a{X}_{i,n})^2\Big) \leq  c_4 n^{-4H(t_0)} \sum_{i,j\in\nu_{n}(t_0)}\big (1+|i-j|\big )^{-2}\nonumber\\
&&\le c_5 n^{-4H(t_0)} n_{t_0} \le  c_6 n^{1-4H(t_0)}v(n),
\end{eqnarray}
where the constants $c_4$, $c_5=2c_4\sum_{l=1}^{\infty} l^{-2}$ and $c_6$ do not depend on $n$. Thus using Markov's inequality, (\ref{var4:}) and (\ref{varsum2}),
we obtain that, for
any $\eta>0$,
\begin{eqnarray}
\label{marcov:}
&&\mathbb P\Big(\Big|\frac{\sum_{i\in\nu_{n}(t_0)}\Delta_a({X}_{i,n})^2}{\sum_{i\in\nu_{n}(t_0)}Var(\Delta_a{X}_{i,n})}-1\Big|>\eta\Big)\leq
\Big(\sum_{i\in\nu_{n}(t_0)}Var(\Delta_a{X}_{i,n})\Big)^{-4}
\eta^{-4}\nonumber\\
&&\hspace{3cm}\times\mathbb E\Big(\sum_{i\in\nu_{n}(t_0)}(\Delta_a{X}_{i,n})^2-\sum_{i\in\nu_{n}(t_0)}Var(\Delta_a{X}_{i,n})\Big)^4\nonumber\\
&&\leq c_{6}'
 \Big(\sum_{i\in\nu_{n}(t_0)}Var(\Delta_a{X}_{i,n})\Big)^{-4}\eta^{-4}v(n)^2n^{2-8H(t_0)}.
\end{eqnarray}
On the other hand, there is a constant $c_7>0$ such that we have for all $n\ge p+1$ and all $i\in\{0,\ldots, n-p-1\}$,
\begin{equation}
\label{mino:var} Var(\Delta_a{X}_{i,n})\geq
c_7n^{-2H(i/n)}.
\end{equation}
Thus, combining (\ref{mino:var}) with (\ref{diffHbis-ant}), it follows that there exists a constant $c_8>0$ such that
for all $n$ big enough at $i\in\nu_{n}(t_0)$,
\begin{equation}
\label{mino:var1} Var(\Delta_a{X}_{i,n})\geq
c_8 n^{-2H(t_0)}.
\end{equation}
Relations (\ref{marcov:}) and (\ref{mino:var1}) imply that, for all $n$ big enough,
\begin{equation}
\label{marcov1}
\mathbb P\Big(\Big|\frac{\sum_{i\in\nu_{n}(t_0)}(\Delta_a{X}_{i,n})^2}{\sum_{i\in\nu_{n}(t_0)}Var(\Delta_a{X}_{i,n})}-1\Big|>\eta\Big)\leq c_2c_6(c_8)^{-4}\eta^{-4}v(n)^{-2}n^{-2}.
\end{equation}
This is equivalent to
$$
\frac{\sum_{i\in\nu_{n}(t_0)}(\Delta_a{X}_{i,n})^2}{\sum_{i\in\nu_{n}(t_0)}Var(\Delta_a{X}_{i,n})}=1+\mathcal O_{\mathbb P}(v(n)^{-2}n^{-2}).
$$
Lemma \ref{iden:X} has been proved. $\square$

 \subsection{Proof of Corollary \ref{lem1:antch5}}
It follows from (\ref{eq4:ant-idenB}) that
\begin{equation}
\label{eq2:lem1:antch5}
Var(\Delta_a{B}_{j,n}^{H(t_0)})=\mathbb E\left\{\Big(\sum_{k=0}^p a_k B_{H(t_0)}\big (\frac{j+k}{n}\big ) \Big)
\Big(\sum_{k'=0}^p a_k B_{H(t_0)}\big (\frac{j+k'}{n}\big ) \Big)\right\}.
\end{equation}
Recall that, for each $s\in [0,1]$,
\begin{equation}
\label{repre2}
B_{H(t_0)}(s)=\int_{\mathbb{R}}\frac{e^{is\xi}-1}{|\xi|^{H(t_0)+1/2}}\ud
\widehat{B}(\xi),
\end{equation}
where $\ud\widehat{B}$ is the Fourier transformation of the white noise
$\ud W$. Relations (\ref{eq2:lem1:antch5}), (\ref{repre2}), (\ref{moment:a:}), the isometry property of the integral $\int_{{\mathbb R}}(\cdot)\,d\widehat{B}$
and Fubini theorem, imply that
\begin{equation}
\label{eq3:lem1:antch5}
Var(\Delta_a{B}_{j,n}^{H(t_0)})= \int_{{\mathbb R}} \frac{\big |\sum_{k=0}^p a_k e^{ik\xi/n}\big|^2}{|\xi|^{2H(t_0)+1}}\ud \xi.
\end{equation}
Finally setting in (\ref{eq3:lem1:antch5}) $\eta=\xi/n$, we obtain the corollary. $\square$

\subsection{Proof of Lemma \ref{lem2:antch5}}
\subsubsection{Proof of (\ref{eq23:ant-idenB})}
Let $\|Y\|_{L^2}=\sqrt{\mathbb E|Y|^2}$ denote the norm of the zero-mean element $Y$ in the space $L^2(\Omega)$ generated by the mBm $\{X(t)\}_{t\in[0,1]}$.  Using the triangle inequality, we obtain
\begin{eqnarray}
\label{eq1:ant-idenB}
&&\|\Delta_a{X}_{j,n}-\Delta_a{B}_{j,n}^{H(t_0)}\|_{L^2}\nonumber\\
&&\le \|\Delta_a{X}_{j,n}-\Delta_a{B}_{j,n}^{H(j/n)}\|_{L^2}
+\|\Delta_a{B}_{j,n}^{H(j/n)}-\Delta_a{B}_{j,n}^{H(t_0)}\|_{L^2},
\end{eqnarray}
where the generalized increments of the fBm $B_{H(j/n)}$, $\Delta_a{B}_{j,n}^{H(j/n)}$,  is defined by
\begin{equation}
\label{eq2:ant-idenB}
\Delta_a{B}_{j,n}^{H(j/n)}=\sum_{k=0}^pa_k{B}_{j+k,n}^{H(j/n)}=\sum_{k=0}^pa_kB_{H(j/n)}\big(\frac{j+k}{n}\big).
\end{equation}
First, let us bound $\|\Delta_a{X}_{j,n}-\Delta_a{B}_{j,n}^{H(j/n)}\|_{L^2}^2=\mathbb E\Big|\Delta_a{X}_{j,n}-\Delta_a{B}_{j,n}^{H(j/n)}\Big|^2$. For each $s\in(0,1)$, using the harmonizable representations  of $X(s)$ (see (\ref{mBm})) and $B_{H(j/n)}(s)$ (see (\ref{repre2})), we write
\begin{eqnarray}
\label{eq5:ant-idenB}
&&\Big|\Delta_a{X}_{j,n}-\Delta_a{B}_{j,n}^{H(j/n)}\Big|=\Big|\sum_{k=0}^pa_k\Big(X\big(\frac{j+k}{n}\big)-B_{H(j/n)}\big(\frac{j+k}{n}\big)\Big)\Big|\nonumber\\
&&=\Big|\sum_{k=0}^pa_k\int_{\mathbb{R}}\Big(\frac{e^{i((j+k)/n)\xi}-1}{|\xi|^{H((j+k)/n)+1/2}}
-\frac{e^{i((j+k)/n)\xi}-1}{|\xi|^{H(j/n)+1/2}}\Big)\ud\widetilde{W}(\xi)\Big|\nonumber\\
&&=\Big|\int_{\mathbb{R}}\Big(\sum_{k=0}^pa_k (e^{i((j+k)/n)\xi}-1)|\xi|^{-1/2}\big( g\big(\frac{j+k}{n},\xi\big)-g\big(\frac{j}{n},\xi\big)\big)\Big)\ud\widetilde{W}(\xi)\Big|,\nonumber\\
\end{eqnarray}
where $g$ is the function defined for each $x\in[0,1]$ and $\xi\in\mathbb{R}\backslash\{0\}$, as,
\begin{equation}
\label{eq6:ant-idenB}
g(x,\xi)=|\xi|^{-H(x)}.
\end{equation}
It follows from (\ref{eq5:ant-idenB}) and the isometry property of the integral $\int_{\mathbb R} (\cdot)\ud\widetilde{W}$ that,
\begin{eqnarray}
\label{eq9:ant-idenB}
&& \mathbb E\Big|\Delta_a{X}_{j,n}-\Delta_a{B}_{j,n}^{H(j/n)}\Big|^2\nonumber\\
&& =\int_{\mathbb{R}}\Big |\sum_{k=0}^pa_k (e^{i((j+k)/n)\xi}-1)|\xi|^{-1/2}\big( g((j+k)/n,\xi)-g(j/n,\xi)\big)\Big |^2 \ud\xi.\nonumber\\
\end{eqnarray}
Observe that
\begin{equation}
\label{eq7:ant-idenB}
\partial_x g(x,\xi)=-H'(x)|\xi|^{-H(x)}\log|\xi|=-H'(x)g(x,\xi)\log|\xi|,
\end{equation}
and
\begin{equation}
\label{eq8:ant-idenB}
\partial_{x}^2 g(x,\xi)=\big(H'(x)^2 \log|\xi|-H''(x)\big) g(x,\xi)\log|\xi|.
\end{equation}
Then (\ref{eq6:ant-idenB}), (\ref{eq7:ant-idenB}), (\ref{eq8:ant-idenB}) and the fact that
\begin{equation}
\label{eq8bis:ant-idenB}
\sup_{x\in [0,1]} \big (|H'(x)|+|H''(x)|\big )<\infty,
\end{equation}
yield that there exists a constant $c_{11}>0$ such that for all $m\in\{0,1,2\}$, $x\in [0,1]$ and $\xi\in{\mathbb R}\setminus\{0\}$,
\begin{equation}
\label{eq7bis:ant-idenB}
\big|\partial_{x}^m g(x,\xi)\big|\le c_{11} \big
(|\xi|^{-H_*}+|\xi|^{-H^*}\big) \big (1+\big|\log |\xi| \big |^m\big).
\end{equation}
where $H_*=\min_{s\in[0,1]}H(s)$ and $H^*=\max_{s\in[0,1]}H(s)$.
Moreover, using for each fixed $\xi\in{\mathbb R}\setminus\{0\}$, a Taylor expansion of order $2$ of
$g(\cdot,\xi)$ at $j/n$, we have, for all $x\in [0,1]$,
\begin{equation}
\label{eq10:ant-idenB}
g(x,\xi)= g(\frac{j}{n},\xi)+(x-\frac{j}{n})\partial_xg(\frac{j}{n},\xi) +(x-\frac{j}{n})^2\int_0^1 (1-\theta) \partial_{x}^2 g(\frac{j}{n}+\theta(x-\frac{j}{n}),\xi)\ud\theta.
\end{equation}
Combining (\ref{eq9:ant-idenB}) with (\ref{eq10:ant-idenB}) we get that
\begin{equation}
\label{eq11:ant-idenB}
\mathbb E\Big|\Delta_a{X}_{j,n}-\Delta_a{B}_{j,n}^{H(j/n)}\Big|^2\le 2 U_{j,n}+2V_{j,n},
\end{equation}
where
\begin{equation}
\label{eq12:ant-idenB}
U_{j,n}=\int_{\mathbb{R}}\Big |\sum_{k=0}^p a_k (e^{i((j+k)/n)\xi}-1)|\xi|^{-1/2} (\frac{k}{n})\partial_xg(\frac{j}{n},\xi)\Big |^2 \ud\xi,
\end{equation}
and
\begin{equation}
\label{eq13:ant-idenB}
V_{j,n} = \int_{\mathbb{R}}\Big |\frac{\sum_{k=0}^p a_k (e^{i((j+k)/n)\xi}-1)}{|\xi|^{1/2}} (\frac{k}{n})^2\int_0^1 (1-\theta) \partial_{x}^2 g(\frac{j}{n}+\theta(x-\frac{j}{n}),\xi)\ud\theta\Big |^2\!\!\!\!\ud\xi.
\end{equation}
Let us now give a suitable bound for $U_{j,n}$. Relations
(\ref{eq12:ant-idenB}), (\ref{moment:a:}) and (\ref{eq7:ant-idenB}) entail that
\begin{eqnarray}
\label{eq14:ant-idenB}
&&U_{j,n}\le   2n^{-2}\int_{\mathbb{R}}\frac{\big |A(\xi/n)\big |^2 \big (H'(j/n)\big)^2\big(\log |\xi|\big )^2}{|\xi|^{2H(j/n)+1}}\ud\xi\nonumber\\
&& \hspace{3cm}+ 2n^{-2} \int_{\mathbb{R}}\frac{\big |A'(\xi/n)\big |^2 \big (H'(j/n)\big)^2\big(\log |\xi|\big )^2}{|\xi|^{2H(j/n)+1}}\ud\xi\nonumber\\
&&\le 2 n^{-2}\int_{\mathbb{R}}\frac{\big(\big |A(\xi/n)\big |^2 + \big |A'(\xi/n)\big |^2\big )\big (H'(j/n)\big)^2\big(\log |\xi|\big )^2}{|\xi|^{2H(j/n)+1}}\ud\xi,
\end{eqnarray}
where $A$ is the trigonometric polynomial defined for $\eta\in{\mathbb R}$ by,
\begin{equation}
\label{eq15:ant-idenB}
A(\eta)=\sum_{k=0}^p a_k e^{ik\eta},
\end{equation}
and $A'$ is its derivative. Observe that all the integrals in (\ref{eq14:ant-idenB}) are finite since
\begin{equation}
\label{eq16:ant-idenB}
\big |A(\eta)\big|=\mathcal O\big (\min\{1,|\eta|\}\big ) \mbox{ and } \big |A'(\eta)\big|=\mathcal O\big (\min\{1,|\eta|\}\big );
\end{equation}
relation (\ref{eq16:ant-idenB}) is in fact a consequence of
(\ref{moment:a:}). Setting in (\ref{eq14:ant-idenB}) $\eta=\xi/n$ and using
(\ref{eq6:ant-idenB}), (\ref{eq7:ant-idenB}), (\ref{eq8bis:ant-idenB}), (\ref{eq7bis:ant-idenB}) and (\ref{diffH}), it
follows that, for all $j\in\nu_n (t_0)$,
\begin{eqnarray}
\label{eq17:ant-idenB}
\nonumber
U_{j,n} &\le & 4 \big (H'(j/n)\big)^2 n^{-2-2H(j/n)}\big (\log n\big)^2\int_{\mathbb{R}}\big(\big |A(\eta)\big |^2 + \big |A'(\eta)\big |^2\big )
\big (g(j/n,\eta)\big )^2\,d\eta\\
\nonumber
&& +4n^{-2-2H(j/n)} \int_{\mathbb{R}}\big(\big |A(\eta)\big |^2 + \big |A'(\eta)\big |^2\big )
\big (\partial_x g(j/n,\eta)\big )^2\,d\eta\\
& \le& c_{12} n^{-2-2H(t_0)}\big (\log n\big )^2,
\end{eqnarray}
where $c_{12}>0$ is a constant non depending on $n$ and $j$. Let us now give a
suitable bound for $V_{j,n}$. Relation (\ref{eq13:ant-idenB}), the triangle
inequality, the inequality $\big (\sum_{k=0}^p b_k\big )^2\le
(p+1)\sum_{k=0}^p b_{k}^2$ for all real numbers $b_0,\ldots, b_p$ and Relation
(\ref{eq7bis:ant-idenB}) imply that,
\begin{eqnarray}
\label{eq18:ant-idenB}
\nonumber
V_{j,n} &\le &(p+1)\sum_{k=0}^p |a_k|^2(k/n)^4\int_{\mathbb{R}}\big |e^{i((j+k)/n)\xi}-1\big |^2 |\xi|^{-1} \nonumber\\
&& \hspace{3cm}\times \Big |\int_0^1 (1-\theta) \partial_{x}^2
g(j/n+\theta(x-j/n),\xi)\ud\theta\Big |^2 \ud\xi\nonumber\\
&\le & c_{11}^2 (p+1)^5 n^{-4}\sum_{k=0}^p
|a_k|^2\int_{\mathbb{R}}\big |e^{i((j+k)/n)\xi}-1\big |^2 |\xi|^{-1}\nonumber\\
&&\hspace{3cm} \times \big (|\xi|^{-H_*}+|\xi|^{-H^*}\big)^2 \big (1+\big|\log
|\xi| \big |^2\big)^2 \ud\xi;
\end{eqnarray}
moreover, we have
\begin{equation}
\label{eq19:ant-idenB}
c_{13}=\sup_{x\in [0,1]} \int_{\mathbb{R}}\big |e^{ix\xi}-1\big |^2 |\xi|^{-1}\big (|\xi|^{-H_*}+|\xi|^{-H^*}\big)^2 \big (1+\big|\log
|\xi| \big |^2\big)^2 \ud\xi<\infty,
\end{equation}
because
$$
x\longmapsto \int_{\mathbb{R}}\big |e^{ix\xi}-1\big |^2 |\xi|^{-1}\big (|\xi|^{-H_*}+|\xi|^{-H^*}\big)^2 \big (1+\big|\log
|\xi| \big |^2\big)^2 \ud\xi,
$$
is a continuous function on the compact interval $[0,1]$. Next, combining
(\ref{eq18:ant-idenB}) with (\ref{eq19:ant-idenB}), we obtain that for all
$n$ big enough and $j\in\nu_n(t_0)$,
\begin{equation}
\label{eq20:ant-idenB}
V_{j,n}\le c_{14} n^{-4},
\end{equation}
where $c_{14}>0$ is a constant non depending on $n$ and $j$. Next, it follows
from (\ref{eq11:ant-idenB}), (\ref{eq17:ant-idenB}) and (\ref{eq20:ant-idenB})
that, for all there is a constant $c_{15}$ not depending on $n$ such that for all $j\in\nu_n(t_0)$,
\begin{equation}
\label{eq21:ant-idenB}
\mathbb E\Big|\Delta_a{X}_{j,n}-\Delta_a{B}_{j,n}^{H(j/n)}\Big|^2
\le c_{15} n^{-2-2H(t_0)}\big (\log n\big)^2.
\end{equation}
Now we can use the same process to bound
$\|\Delta_a{B}_{j,n}^{H(j/n)}-\Delta_a{B}_{j,n}^{H(t_0)}\|_{L^2}^2$ as we did for bounding $\|\Delta_a{X}_{j,n}-\Delta_a{B}_{j,n}^{H(j/n)}\|_{L^2}^2$. It suffices to point out that the term $k/n=\mathcal O(n^{-1})$ in $U_{j,n}$ (in (\ref{eq12:ant-idenB})) and $V_{j,n}$ (in (\ref{eq13:ant-idenB})) are replaced with $|t_0-j/n|=\mathcal O(v(n))$ in this case. Hence It turns out that  there is a constant $c_{16}>0$ not depending on $n$ such that for $j\in\nu_{n}(t_0)$,
\begin{equation}
\label{eq22bis:ant-idenB}
\mathbb E\Big|\Delta_a{B}_{j,n}^{H(j/n)}-\Delta_a{B}_{j,n}^{H(t_0)}\Big|^2
\le c_{16}\big(v(n)\log(n)\big)^2 n^{-2H(t_0)}.
\end{equation}
Next, putting together (\ref{eq1:ant-idenB}), (\ref{eq21:ant-idenB}),
(\ref{eq22bis:ant-idenB}) and the inequality $v(n)\ge n^{-1}$, we get (\ref{eq23:ant-idenB}). $\square$

\subsubsection{Proof of (\ref{ident1})}
First by using the fact that $Var(\Delta_a{B}_{j,n}^{H(t_0)})=\|\Delta_a{X}_{j,n}\|_{L^2}^2$ and the triangle inequality we can write
\begin{eqnarray*}
&&\left|\frac{Var(\Delta_a{X}_{j,n})}{Var(\Delta_a{B}_{j,n}^{H(t_0)})}-1\right|=\frac{\big|\|\Delta_a{X}_{j,n}\|_{L^2}^2-\|\Delta_a{B}_{j,n}^{H(t_0)}\|_{L^2}^2\big|}{\|\Delta_a{B}_{j,n}^{H(t_0)}\|_{L^2}^2}\\
&&=\frac{\big|\|\Delta_a{X}_{j,n}\|_{L^2}-\|\Delta_a{B}_{j,n}^{H(t_0)}\|_{L^2}\big|\big(\|\Delta_a{X}_{j,n}\|_{L^2}+\|\Delta_a{B}_{j,n}^{H(t_0)}\|_{L^2}\big)
}{\|\Delta_a{B}_{j,n}^{H(t_0)}\|_{L^2}^2}\\
&&\leq
\frac{\|\Delta_a{X}_{j,n}-\Delta_a{B}_{i,n}^{H(t_0)}\|_{L^2}\big(\|\Delta_a{X}_{j,n}\|_{L^2}+\|\Delta_a{B}_{j,n}^{H(t_0)}\|_{L^2}\big)
}{\|\Delta_a{B}_{j,n}^{H(t_0)}\|_{L^2}^2}.
\end{eqnarray*}
 Then (\ref{ident1}) follows from Corollary~\ref{lem1:antch5}, (\ref{eq23:ant-idenB}) and Lemma~\ref{rem1:antch5}. $\square$

\subsubsection{Proof of (\ref{eq1:lem3:antch5})} 
Using the triangle inequality and (\ref{ident1}) we obtain,
\begin{eqnarray*}
&& \bigg|\frac{\sum_{j\in\nu_n(t_0)}Var(\Delta_a{X}_{j,n})}{\sum_{j\in\nu_n(t_0)}Var(\Delta_a{B}_{j,n}^{H(t_0)})}-1\bigg|\\
&&=\bigg|\frac{\sum_{j\in\nu_n(t_0)}Var(\Delta_a{X}_{j,n})-\sum_{j\in\nu_n(t_0)}Var(\Delta_a{B}_{j,n}^{H(t_0)})}{\sum_{j\in\nu_n(t_0)}Var(\Delta_a{B}_{j,n}^{H(t_0)})}\bigg|\\
&&\le\frac{\sum_{j\in\nu_n(t_0)}Var(\Delta_a{B}_{j,n}^{H(t_0)})\bigg|\frac{Var(\Delta_a{X}_{j,n})}{Var(\Delta_a{B}_{j,n}^{H(t_0)})}-1\bigg|}{\sum_{j\in\nu_n(t_0)}Var(\Delta_a{B}_{j,n}^{H(t_0)})}\\
&&\le c\log(n)v(n),
\end{eqnarray*}
where $c>0$ is a constant which does not depend on $n$. $\square$

\subsection{Figures for simulation study}
\label{Appendix:figures}
In this subsection, we provide more evidences on convergence behavior of the PHE estimation comparison for functional forms, using additional functional forms that are not limited to what we have shown in Figure \ref{fig::sim_t2_ep}, Figure \ref{fig::sim_t2_ep} and Figure \ref{fig::sim_t3_ep}. The specific functional forms that we present in this section contain:

\begin{itemize}
\item $\Phi(t,X(t)) = X(t)$;
\item $\Phi(t,X(t)) = X(t)^2$;
\item $\Phi(t,X(t)) = \sin(t)^2 + X(t)^2$;
\item $\Phi(W(t),X(t)) = W(t)^2 + X(t)^2$;
\end{itemize}

\begin{center}
\begin{figure}[ph]
\centering
\includegraphics[scale = 0.8]{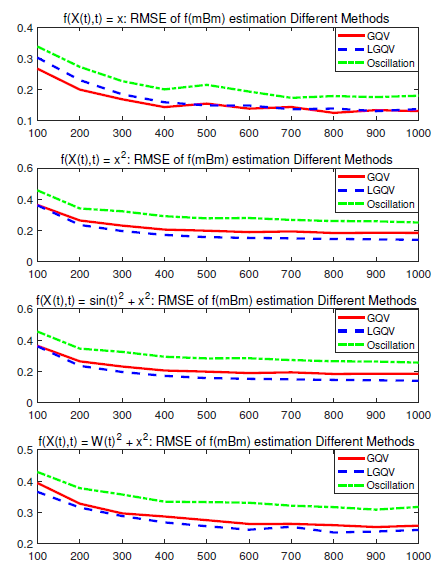}
\newline
\caption{\normalsize The four graphs compare the RMSE of the classic GQV, LGQV and oscillation methods over 100 simulations for functional forms: 1. $\Phi(t,X(t))=X(t)$; 2. $\Phi(t,X(t))=X(t)^2$; 3. $\Phi(t,X(t))=\sin(t)^2+X(t)^2$; 4. $\Phi(W(t),X(t))=W(t)^2+X(t)^2$, respectively.}
\label{fig::four_rest_plots}
\end{figure}
\end{center}

Figure \ref{fig::four_rest_plots} illustrates the RMSE comparison between PHE estimation using LGQV and using benchmark methods (GQV method and Oscillation method) these four additional functional forms, in supplement to the comparisons in Section \ref{sec:sim}. 

For the case when $\Phi(t, X(t)) = X(t)$, the observed mBm returns to itself (no additional functional transformation). We find that classic GQV estimation outperforms the LGQV method for most of the time. However, for the functional of mBm series (from second case to the fourth case above), LGQV outperforms both benchmark methods in terms of its lower  RMSE.  

\newpage
\bibliographystyle{plain}
\bibliography{qf}

\end{document}